\documentclass[useAMS,usenatbib]{mn2e}
\usepackage{graphicx}
\usepackage{latexsym}
\usepackage{xspace}

\title[The non-spherical nature of collapsing cores]{A quantification of the non-spherical geometry and accretion of collapsing cores}
\author[Smith et~al.]{Rowan J. Smith$^{1}$\thanks{Email: rowan@ita.uni-heidelberg.de}, Simon C. O. Glover$^{1}$, Ian A. Bonnell$^{2}$, Paul C. Clark$^{1}$, Ralf S. Klessen$^{1,3}$ \\
$^1$ Zentrum f\"ur Astronomie der Universit\"at Heidelberg, Institut f\"ur Theoretische Astrophysik, Albert-Ueberle-Str. 2, 69120 Heidelberg, Germany \\
$^2$ SUPA, School of Physics \& Astronomy, University of St Andrews, North Haugh, St Andrews, Fife, KY16 9SS, UK \\
$^3$ Kavli Institute for Particle Astrophysics and Cosmology, Stanford University, Menlo Park, CA 94025, USA\\
 }
 
\begin{document}

\pagerange{\pageref{firstpage}--\pageref{lastpage}} \pubyear{2010}

\maketitle

\label{firstpage}

\def\mnras{MNRAS}
\def\apj{ApJ}
\def\aap{A\&A}
\def\apjl{ApJL}
\def\apjs{ApJS}
\def\bain{BAIN}
\def\pasp{PASP}
\def\araa{ARA\&A}
\def\ga{\sim}
\def\nat{Nature}
\def\aj{AJ}
\def\pasj{PASJ}


\newcommand{\eq}{Equation }
\newcommand{\fig}{Figure }
\newcommand{\msun}{\,M$_{\odot}$ }
\newcommand{\gcmc}{\,g\,cm$^{-3}$}
\newcommand{\kms}{\,kms$^{-1}$}
\newcommand{\tab}{Table }
\newcommand{\gcms}{\,g\,cm$^{-2}$\xspace}
\newcommand{\E}{\times 10}

\begin{abstract}
We present the first detailed classification of the structures of Class 0 cores in a high resolution simulation of a giant molecular cloud. The simulated cloud contains 10$^{4}$ \msun and produces over 350 cores which allows for meaningful statistics. Cores are classified into three types according to how much they depart from spherical symmetry. We find that three quarters of the cores are better described as irregular filaments than as spheres. Recent Herschel results have shown that cores are formed within a network of filaments, which we find has had a significant impact on the resulting core geometries. We show that the column densities and ram pressure seen by the protostar are not uniform and generally peak along the axes of the filament. The angular momentum vector of the material in the cores varies both in magnitude and direction, which will cause the rotation vector of the central source to fluctuate during the collapse of the core. In the case of the more massive stars, accretion from the environment outside the original core volume is even more important than that from the core itself. This additional gas is primarily accreted onto the cores along the dense filaments in which the cores are embedded, and the sections of the surfaces of the cores which do not coincide with a filament have very little additional material passing through them. The assumption of spherical symmetry cannot be applied to the majority of collapsing cores, and is never a good description of how stars accrete gas from outside the original core radius. This has ramifications for our understanding of collapsing cores, in particular their line profiles, the effect of radiation upon them and their ability to fragment.
\end{abstract}


\section{Introduction}
Understanding the birth of stars from their parent molecular clouds has been a longstanding problem in astrophysics, and over the years it has received considerable attention \citep[e.g.][]{MacLow04,McKee07}. While early investigations used spherical models for the collapse of gas to stellar densities \citep[e.g.][]{Larson69,Penston69,Shu77}, it has been known for some time that structure within molecular clouds is both irregular and filamentary \citep{Blitz93}. 

It is observationally well established that molecular clouds are filamentary in nature \citep[e.g.][]{Schneider79,Low84,Bally87,Loren89,Johnstone99}. This has been further emphasised in recent results from the Herschel space telescope which have revealed that cores are formed within a network of narrow filaments \citep{Menshchikov10,Andre10,Henning10,Miville-Deschenes10,Ward-Thompson10}. Additionally, the formation of irregular structures has been a long-standing feature in simulations of turbulent, dynamic star formation. For example, in simulations with turbulence and self-gravity \citep{Klessen01,Vazquez-Semadeni05,Jappsen05,Heitsch08b,Bate09,Walch09,Federrath10} filaments are a ubiquitous phenomena. Indeed, \citet{Burkert04} have shown that filaments are a natural consequence of gravitational focussing during the collapse of any mass distribution which departs from circular symmetry. 

Filaments are prone to fragmentation and several authors have analysed their stability \citep{Ostriker64,Inutsuka92,Hennebelle03}. \citet{Inutsuka97} found that quasi-equilibrium filaments fragment into dense cores separated by about four times the filament diameter. Therefore filaments can be thought of as the birthing grounds of pre-stellar cores. This raises the question: how much of this initial structure is retained when the core is formed, and how does this affect accretion onto the core?

Observationally, there have been many surveys which have aimed to describe the structure of dense cores at scales of $0.1$ pc or larger \citep{Myers91,Bacmann00,Jones01,Jones02,Goodwin02,Tassis07,Tassis09}. These studies have found that the models which best described cores were extended or triaxial structures, rather than spheres. Theoretical studies of the same topic have found similar results \citep[e.g.][]{Ballesteros-Paredes03,Tilley04,Li04,Klessen05}. Recent observations of isolated cores using mid-infrared shadows by \citet{Stutz09} demonstrate that these objects have complex morphologies.  However, perhaps the most interesting scale to probe is at around $1000$~AU, where the accretion onto the central proto-star and the formation of a disk will occur. Recently \citet[][hereafter T10]{Tobin10} published observations in this regime. These authors studied the collapsing envelopes of $22$ Class 0 protostars using Spitzer IRAC observations from nearby star-forming clouds and found that highly non-axisymmetric geometries were common. These asymmetries do not appear to be caused by outflows, as the observed outflow cavities are narrow, and appear spatially distinct from the dense structures. Previous large scale simulations were unable to resolve structure on these small scales; this work represents the first quantification of simulated core geometries on the scales at which the proto-star is actually forming.

There are significant consequences if departures from spherical symmetry are common. Infalling cores are typically identified by the observation of blue-skewed double-peaked line profiles \citep{Evans99}. This is the line profile expected from a spherical core undergoing collapse, but if there are asymmetries in the core, then the profile may be distorted along some viewing angles. Thus, if a significant population of cores are not spherical, then we may be underestimating the number of cores undergoing dynamic collapse. The collapse of a spherically symmetric distribution leads naturally to a single central condensation. However, a more complex geometry is likely to lead to further fragmentation which has implications for the binary population \citep[e.g.][]{Bonnell92,Burkert93} formed from the cores. In studies of massive star formation, spherical symmetry increases the difficulty of overcoming radiation pressure during the accretion process \citep{Kahn74}. However, if there are under-dense regions surrounding the protostar from its birth, this obstacle is more easily overcome. Radiation will also penetrate more easily into clumpy, non-axisymmetric cores than into uniform spherical cores of the same mass, leading to higher gas and dust temperatures and higher fractional ionizations within the cores \citep{Padoan04b,Bethell04,Bethell07}, all of which will potentially influence the dynamics of core collapse.

Beyond the collapse of the initial core, accretion from the environment may play an important role in the formation of stellar systems, particularly in the case of massive stars, as shown in \citet*[][hereafter SLB09]{Smith09b}. The geometry of the flow of accreted material from the environment onto the core has important implications for both the rate at which additional gas can be accreted, and for how long the accretion can continue. For instance, gas can be transferred onto the core more rapidly if it is being accreted from all angles compared to just one direction. \citet{Myers08} argues that runaway accretion should only be possible when the gas surrounding the core is extended in all three dimensions. However, if gas approaches the core along a few very dense flows, it may be more difficult to shut off inflow by radiative processes \citep[e.g.][a,b]{Dale08,Krumholz09,Peters10}, and mechanisms such as jets may need to be invoked to finally stop accretion \citep{Wang10}.


The aims of this paper are therefore twofold. Firstly we aim to quantify how common extreme anisotropies are in the Class 0 cores formed within a high-resolution simulation of a giant molecular cloud (GMC). This should provide robust statistics which can be compared to observations and provide initial conditions for future modeling of core collapse. Further, this paper aims to trace how material originally external to the core approaches the forming proto-star and to determine whether accretion onto a core takes place in an inhomogeneous manner.

\section{Method} 

\subsection{The Simulation}
This paper is the latest in a series of papers, \citet{Smith09} (hereafter SCB09) and SLB09, that have utilised large simulations of a giant molecular cloud to study in depth the properties of clustered prestellar cores, and the simulation here is the same as that described in SLB09. The smoothed particle hydrodynamics (SPH) method is used to follow the evolution of a $10^{4}$ M$_{\odot}$ cloud over $1.02$ free-fall times or $\approx 6.6 \times 10^{5}$ years. The cloud resembles a long filament, with a length of $10$ pc and a radius of $3$ pc. The cloud has a local density gradient such that one end of the cloud is gravitationally bound
while the other end of the cloud is unbound, and thus a range of star-forming environments are present. The gas has internal decaying turbulence following a Larson-type $P(k) \sim k^{-4}$ power law that is normalised such that the total kinetic energy balances the total gravitational energy in the cloud at the start of the simulation. Sink particles are used to model sites of star formation \citep{Bate95}, and have an outer accretion radius of $10^{-3}$ pc and an inner accretion radius of $10^{-4}$ pc. At the outer accretion radius the gas will be accreted by the sink if it is unequivocally bound to it. Any remaining gas will only be accreted if it reaches the inner accretion radius.

The simulation has $15.5$ million SPH particles on two levels to maximise the numerical resolution in the cores. Regions that underwent star formation and formed sink particles or were subsequently accreted by these sink particles were identified in a lower resolution simulation. These particles were then replaced in the initial conditions with 9 lower-mass particles, conserving the mass and kinetic energy of the initial conditions, yielding a maximum mass resolution of $0.0167 \: {\rm M_{\odot}}$. 
This high resolution is essential in order to sufficiently resolve the cores to study their geometry. The simulation was re-run from the initial conditions with the locally increased mass resolution. 

The simulation treats the thermal content of the cloud through a barotropic equation of state
\citep{Larson05,Jappsen05} and a thermal heating contribution is included to represent radiative feedback from the sinks. The heating from the newly formed stars  is approximated by way of a grid of previously computed Monte Carlo radiative transfer models of young stars \citep{Robitaille06}. A one-dimensional temperature profile was derived from the youngest of these models as a function of stellar mass and distance which gives a very rough estimate of the radiative feedback. If anything, this procedure should overestimate the gas temperatures. Nevertheless, it gives an estimate of the emission expected in regions of massive star formation. From the Monte Carlo models, the temperature at a distance $r$ due to the radiative feedback is given by
\begin{equation}
T(r) = \left \{ \begin{array}{lr}
100 \left(\frac{m}{ 10 \: {\rm M}_{\odot}}\right)^{0.35} \left(\frac{r}{1000 {\rm AU}}\right)^{-0.45}  \: {\rm K} &  m \leq 10 M_{\odot}, \\
100 \left(\frac{m}{ 10 \: {\rm M}_{\odot}}\right)^{1.11} \left(\frac{r}{1000 {\rm AU}}\right)^{-0.5}  \: {\rm K} &  m > 10 M_{\odot},\\
\end{array} \right.
\end{equation}
where $m$ is the mass of the protostar (here assumed equal to the sink particle mass). The gas temperature around the young stars is set to be the maximum of the temperature from either the barotropic equation of state or the radiative feedback. This ensures a maximal effect from the radiation.

\subsection{Column density surface projections}
To study the local environment of the proto-stars, we make maps of the column density of the core of gas surrounding the sink. These are made from the perspective of the central proto-star looking outward. We define the core to be all material within $R_c = 0.01$ pc of the central sink.  In order to ensure that our core sample has a constant volume we do not use a clumpfinding approach. This ensures that differences in the column densities are due to internal density enhancements, and not due to differing radial extents. In SCB09 it was found that the typical radius within which the bound material of a pre-stellar core was contained was  $r=1.16\E^{-2}$ pc. Therefore our selected size of $0.01$ pc represents the volume which will first collapse. Although the Class 0 classification scheme is based on low mass YSOs, we do form high mass stars in this simulation. However, as shown in SLB09 there are no high mass pre-stellar cores in this simulation. Therefore we use the term `Class 0' to refer to all of our collapsing cores.

We use Hammer projections to show the column density surface. In this projection, a point at $(\lambda,\phi)$ on a spherical surface will correspond to position $(x,y)$ on the projected flat surface, according to the transformation;
\begin{equation}\label{Hlong}
\lambda= 2 \arctan\left[\frac{zx}{2(2z^2-1)}\right]
\end{equation}
\begin{equation}\label{Hlatt}
\phi=\arcsin(zy)
\end {equation}
where
\begin{equation}\label{Hz}
z=\sqrt{1-\left(\frac{1}{4}x\right)^2 - \left(\frac{1}{2}y\right)^2}.
\end{equation}
This projection is not conformal, but area is conserved, which is essential for the following calculation of column density.

The first stage to make the  Hammer projections is to find the particles within the core volume for every sink just after it has formed. Cores that are within $2 R_c$ of an existing sink are rejected. In addition, to ensure that we have sufficient resolution, we only use cores with over $500$ SPH particles. This is $10$ times higher than our minimum resolution of $50$ particles. In practice, the cores generally have $\sim1500$ particles

We select lines of sight from a uniform ($50\times100$) grid on the Hammer surface, in order to ensure equal area sampling. This equates to $3895$ lines of sight within the projected elliptical surface. We calculate the density at each point along the line of sight using the standard SPH density equation,
\begin{equation}\label{sphdens}
\rho=\sum_{j=1}^{N}m_{j}W(\mathbf{r}-\mathbf{r_{j}},h).
\end{equation}
where $m_{j}$ is the mass of the $j$'th particle, $W(\mathbf{r}-\mathbf{r_{j}},h)$ is the smoothing kernel and $N$ is the number of particles within a smoothing length, $h$. The densities are then integrated along the line of sight to obtain the column density. A ram pressure map of the core can be made in a similar manner, by integrating the product of density and velocity along the line of sight.

\subsection{Angular covariance}
To categorise the column density surfaces, the angular covariance of the high density gas is calculated for each core projection. The covariance is
\begin{equation}\label{covariance}
cov(d\theta) = Q(d\theta)-\bar{N}^2,
\end{equation}
where $\bar{N}$ is the mean column density and
\begin{equation}\label{corrfunction}
Q(d\theta) = \frac{\sum_{1}^{n} q_{ij}(d\theta_{ij})} {n},
\end{equation}
where $q_{ij}=N_i*N_j$ is the product of the column density, $N$, at points $i$ and $j$ which have an angular separation in the range $d\theta$, and $n$ is the number of pairs $ij$ separated by $d\theta$. 
The covariance is calculated at five degree intervals, as this allows each bin to be well filled while still sampling the angular range finely enough to see small-scale structure.

If the column density distribution is uniform, then we expect $cov(d\theta)$ to be zero everywhere as the column densities will be equal to the average. However, if there is structure in the column density map
with a particular angular scale, for example in the form of ten degree wide blobs, then the value of $cov(d\theta)$ will be higher at this angular separation as the structure is highly correlated on this scale.

\section{The Structure of Protostellar Cores}\label{sec:stcores}

\subsection{Surface Projections}

\begin{figure*}
\begin{center}
\includegraphics[width=6in]{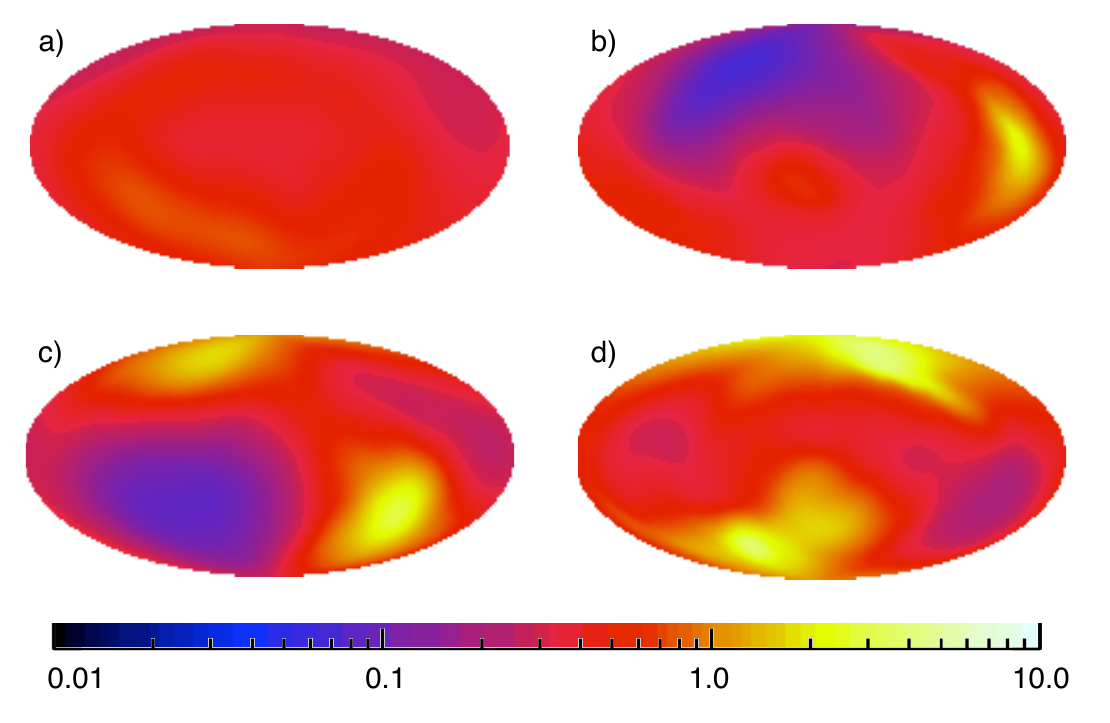}
\caption{The column density projections of four representative cores in \gcms. The column densities are not uniform and are consistent with filamentary geometries.}
\label{CDprojections}
\end{center}
\end{figure*}

The column density projections of four representative cores are shown in \fig \ref{CDprojections}. Core (a) has a fairly uniform column density. However, the other three cores have at least one preferred direction where the column density is higher.
This distribution is consistent with the high density material being distributed along a filament. If the sink particle is situated at the end of the filament, then the column density is higher in the direction of the filament, as in core (b). If the sink particle is located within the filament, then the column density peaks where the filament intersects the core volume, and falls off perpendicular to the filament. Cores (c) and (d) represent this situation. Core (c) has a filament which is denser on one side than the other, and core (d) has a filament which is equally dense on both sides,  showing that the sink is located at the centre of the filament.

The column density seen by the central sink is almost entirely due to the core itself. \fig \ref{CDprojections} is calculated using only the material within $0.01$ pc of the central sink, but when this range was extended to $1.0$ pc, there was no significant change in the column density projection. This is because the density of the gas close to the sink is so high, and covers such a large angle, that it completely dominates the column density.

\begin{figure*}
\begin{center}
\begin{tabular}{c c}
\includegraphics[width=2.5in]{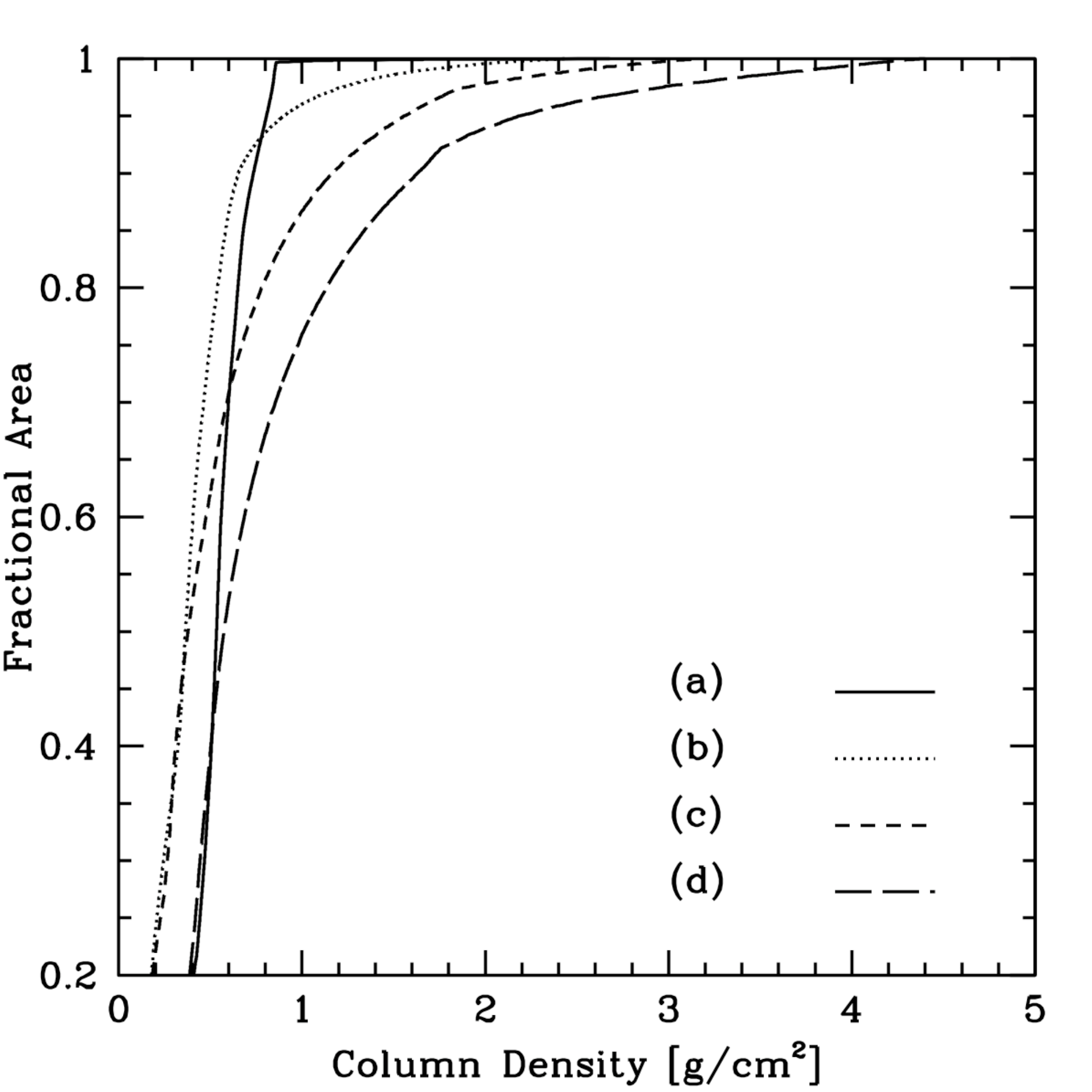}
\includegraphics[width=2.5in]{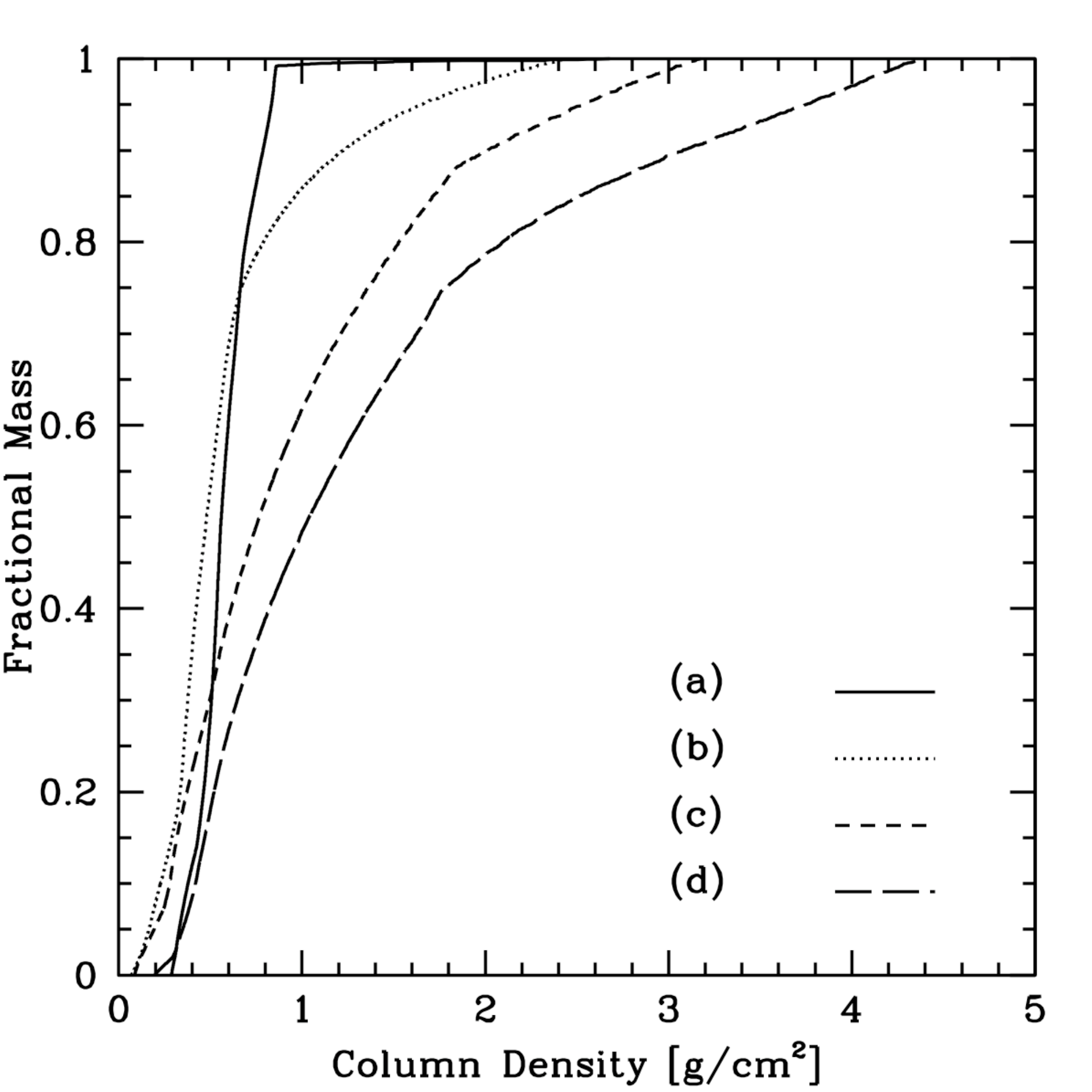}\\
\end{tabular}
\caption{The fraction of the area and mass below a given column density for the four cores shown in \fig \ref{CDprojections}. The non-uniform column density distributions mean that a significant fraction of the mass is at high column densities compared to the surface area coverage.}
\label{CDarea}
\end{center}
\end{figure*}

The range of column densities seen by the central sink is shown in \fig \ref{CDarea} as a percentage of the total core surface area and mass.
For the symmetric core (a) nearly all of its area and mass is at column densities below $1$ \gcms. However, the column densities reach higher values where there are filaments. The spatial clustering in the column density projection means that a larger proportion of the mass is at higher column densities than a simple average may suggest. \tab \ref{area_table} shows the percentage of the area and mass of the cores that is above a column density of $1$ \gcms. In the case of core (d), more than half of its mass is above $1$ \gcms, but this is contained within only $24\%$ of the core's surface. This has important implications for the action of radiative feedback on the core: a large fraction of the core mass will be shielded by a high extinction, and hence will be comparatively unaffected by any feedback, while a large fraction of the surface of the core will be covered by low extinction material, allowing for the easy escape of radiation along these low extinction sightlines. Similar arguments were made in \citet{Dale08} regarding the effects of ionising winds on a star cluster. As the ram pressure seen by the central sink is simply the product of the column density with the gas radial velocity, the pressure towards the centre of the core is also highly non-uniform.

\begin{table}
	\centering
	\caption{The total area and mass above a column density of $1$ \gcms in the cores shown in \fig \ref{CDprojections}. In the filamentary cores a significant fraction of the mass is shielded by high column densities.}
		\begin{tabular}{c c c}
   	         \hline
	         \hline
	         Core & Area & Mass\\
	         & above $1$ \gcms& above $1$ \gcms \\
	         \hline
	         a & $0.2\%$ &$0.6\%$ \\
	         b & $4\%$ & $14\%$ \\
	         c & $13\%$ & $38\%$ \\
	         d & $24\%$& $52\%$ \\
		\hline
		\end{tabular}
	\label{area_table}
\end{table}

\subsection{Global Classifications}\label{sec:global}
\begin{figure}
\begin{center}
\includegraphics[width=2.5in]{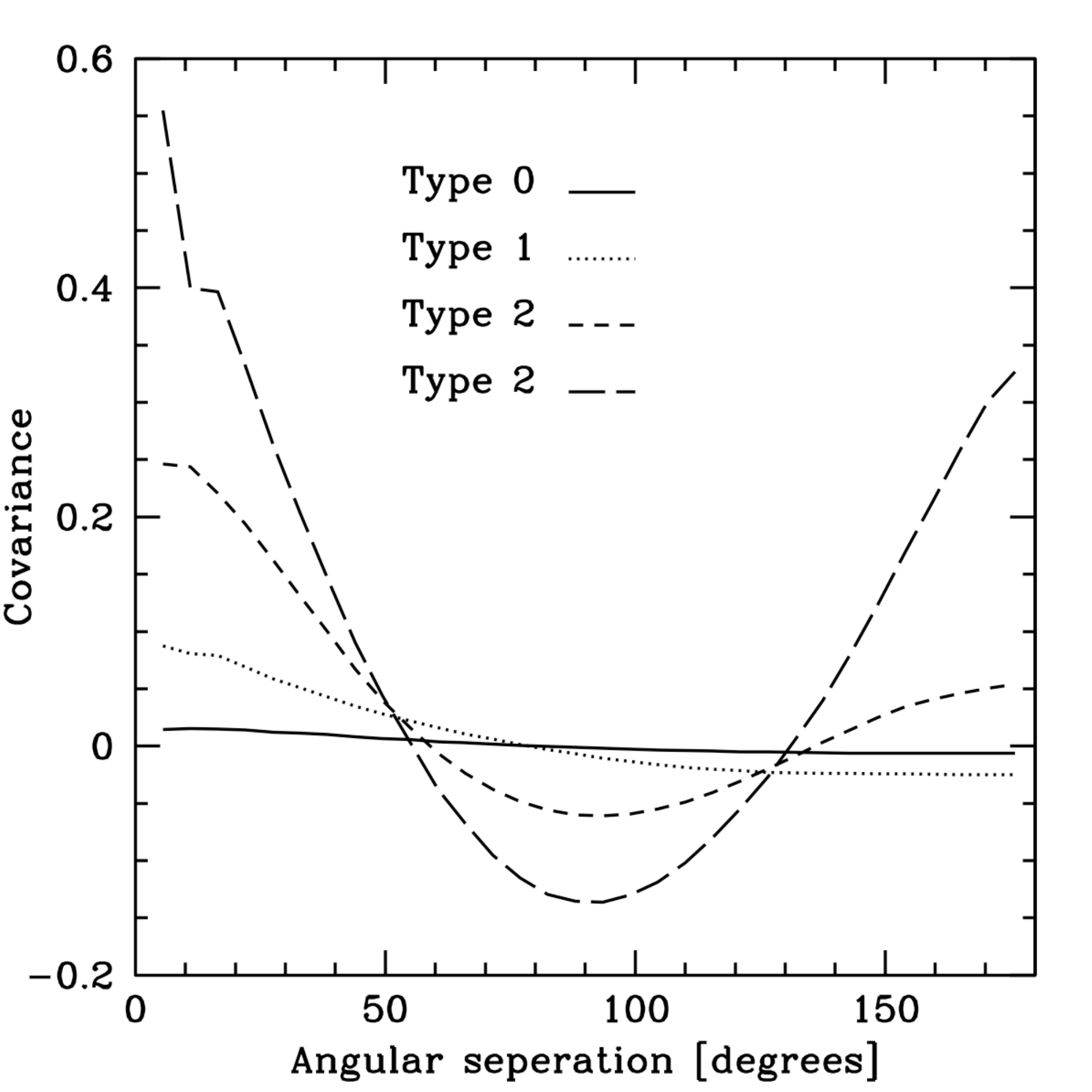}
\caption{Covariance functions for the four representative cores displayed earlier. The key shows 
the classification of each core. Type 0 cores have a relatively flat covariance function with no large over-densities. Type 1 cores have a decreasing covariance function, and have one large over-density. Type 2 cores have a double-peaked covariance function, and have two large over-densities. }
\label{Corr_types}
\end{center}
\end{figure}

That star formation often takes place in filaments is not in itself a new result, although the impact this has on the environment `seen' by the proto-stars is perhaps surprising. However, due to the large number of stars formed in a variety of environments within this simulation, it is possible to quantify how common this is. Obviously it is impractical to show all the column density projections, so the cores are classified according to the shapes of their angular covariance functions. \fig \ref{Corr_types} shows the covariance functions of the four cores shown in \fig \ref{CDprojections}. These covariance functions are typical of the entire data-set and using these shapes it is possible to classify the cores into three types according to the number of over-densities they contain. \citet{Tobin10} have carried out a similar classification on their Class 0 sources, and have independently proposed a very similar scheme.

However before discussing the differences, it is important to note that all the covariance functions are similar in that the greatest covariance is always seen at small separations. This reflects the fact that no column density map was completely uniform, and that in every case high column densities were concentrated at a few centrally peaked locations, rather than distributed randomly throughout the core surface.

Type 0 cores have no strong over-densities, and therefore have a covariance function which is close to $0$ at all angular seperations. Core (a) in \fig \ref{CDprojections} corresponds to this type. We define Type 0 cores as those whose maximum or minimum value departs by less than $0.01$ from zero. They have no strong over-densities, and correspond to the cores in \fig 3 of T10. 

Type 1 cores have a steeper fall off in their covariance function, and the covariance does not become greater than zero again at large angular scales. Core (b) is an example of this type. Physically, this type of covariance function represents a core with one strong over-density. Type 1 cores correspond to the cores in \fig 2 of T10. 

Type 2 cores also have a steep fall off in their covariance function. However, at larger angular separations the covariance becomes greater than zero again. This reflects the fact that in these cores there is more than one strong accretion stream. Moreover the covariance function has a minimum at $90^\circ$ and a second maximum at $180^\circ$, which strongly suggests a filamentary geometry. Cores (c) and (d) are both examples of this type. Type 2 cores correspond to the cores in \fig 1 of T10.

\tab \ref{type_table} shows the number of cores assigned to each classification and their percentage of the total number. Three quarters of the cores are dominated by an irregular filament, and therefore this geometry represents the most common initial state of a collapsing core in the molecular cloud. It is clear that in the majority of cases, the dense cores still strongly retain the geometry of the filament in which they are formed. This can be understood as during the collapse of the filaments there is no internal pressure support that can slow the collapse sufficiently for a spherical core to form.

\begin{table}
	\centering
	\caption{The number of cores assigned to each classification at the end of the simulation. Three quarters of the cores are filamentary (Type 1 or 2).}
		\begin{tabular}{c c c}
   	         \hline
	         \hline
	         Type & Number & Percentage \\
	         \hline
	         0 = spherical & $86$ & $24.2\%$\\
	         1 = 1 sided & $120$ & $33.7\%$\\
	         2 = 2 sided & $150$ & $42.1\%$\\
		\hline
		\end{tabular}
	\label{type_table}
\end{table}

The cores have distinct differences in their mass and column densities, as shown in \tab \ref{type_props}. The filaments concentrate more mass in the cores, and allow them to have a greater range in column density across their structure. The cores which are in the middle of a filament have the largest masses and column densities. The distribution of column densities is highly skewed, as shown earlier in \fig \ref{CDarea}, and a high density tail of column densities means that the range in column densities is larger than the average itself.

\begin{table}
	\centering
	\caption{The average masses and column densities of the cores.}
		\begin{tabular}{c c c c}
   	         \hline
	         \hline
	         Type & Avg. Mass & Avg. Density & Avg. Density Range \\
	         & \msun &\gcms &\gcms\\
	         \hline
	         0 & $0.429$ & $0.427$ & $0.605$\\
	         1 & $0.946$ & $0.713$ &$3.246$\\
	         2 & $1.594$ & $1.108$ & $7.712$\\
		\hline
		\end{tabular}
	\label{type_props}
\end{table}

\subsection{Radial Profiles}
\begin{figure*}
\begin{center}
\begin{tabular}{c c}
\includegraphics[width=2.5in]{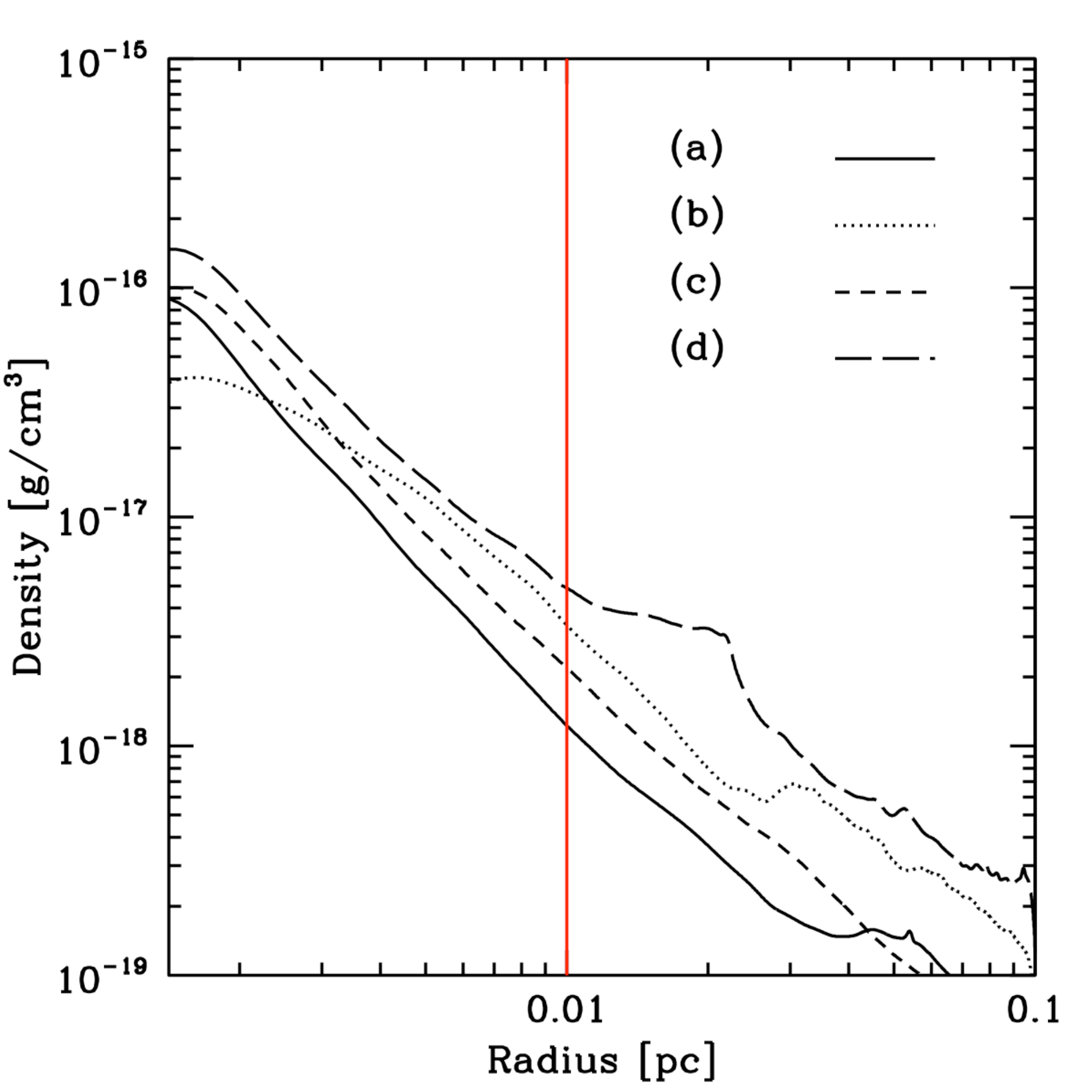}
\includegraphics[width=2.5in]{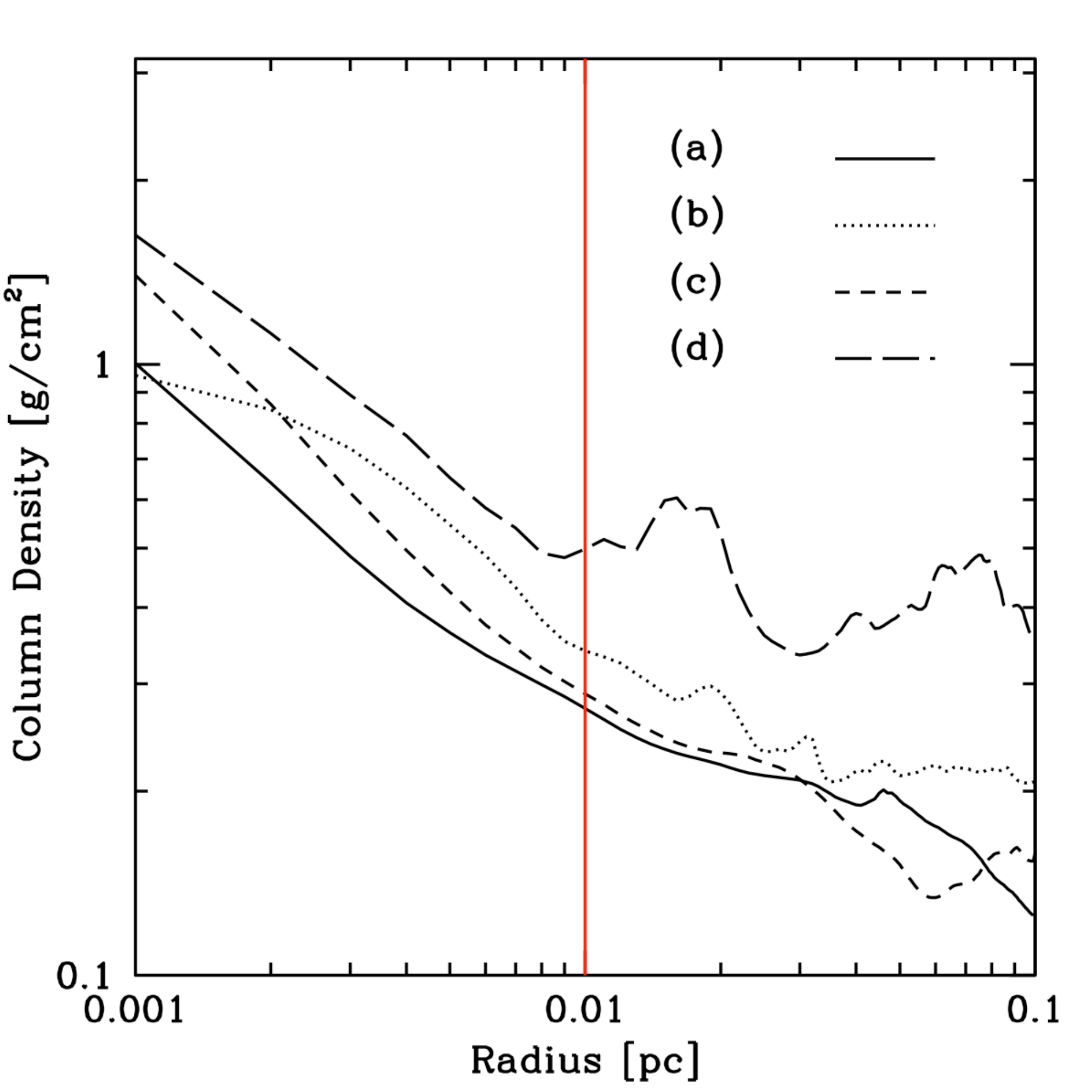}\\
\end{tabular}
\caption{Volume-averaged radial density profiles \textit{(left)} and projected column density profiles along an arbitrary viewing angle \textit{(right)} for the four representative cores. The red vertical line shows our fiducial core radius $R_c=0.01$ pc. Note that the column density profiles will change depending on the viewing angle.}
\label{radial}
\end{center}
\end{figure*}

While the core surface projections are highly anisotropic, the averaged radial profiles are still consistent with previous work since as the cores are collapsing objects, they are centrally condensed, and their density profile decreases outward according to a power law. To confirm this we fitted the power law $\rho\propto r^{-n}$ to the volume averaged density profiles and obtain values for the exponent $n$ for each core. \tab \ref{slopes_table} shows the average power law exponent, $n$, fitted to the radial profiles of the entire core population in various ranges. Within the core volume ($r<R_c$), the volume averaged profile has an exponent of $n=2.04 \pm 0.38$, consistent with the $n=2$ that is expected from the dynamic collapse of an isothermal sphere \citep{Larson69,Penston69}. \fig \ref{radial} shows the volume weighted density profiles and the column density projections for the four representative cores. As shown by \citet{Ballesteros-Paredes03} and \citet{Hartmann04} core density profiles are artificially smoothed by two effects. Firstly, by taking an azimuthal angle average as shown in the volume averaged density profiles of \fig \ref{radial}, and secondly by projection effects as shown in the column density projections in the right hand panel. Thus a smooth power law is obtained from complex structure.

\begin{table}
	\centering
	\caption{The mean power law exponent $n$ of the volume-averaged radial density profiles of the cores, fitted over two different ranges of radii. The standard deviation of $n$ is also listed.}
		\begin{tabular}{c c c}
   	         \hline
	         \hline
	         Fitted range & $n$ & $\sigma$ \\
	         \hline
	         r $<0.01$pc & $2.04$ &$0.38$ \\
	         r $<0.1$pc & $1.57$ &$0.26$ \\
		\hline
		\end{tabular}
	\label{slopes_table}
\end{table}

Observationally,  it is extremely hard to probe the $r=0.01$~pc scales just discussed, and so we also show the slopes fitted to radii of $r \leq 0.1$~pc in \tab \ref{slopes_table}. The mean power law exponent in this case is $n=1.57 \pm 0.26$, which is close to the $n=1.5$ value expected for a freely falling envelope \citep{Young03}. \citet{Shirley02} and \citet{Young03} carried out radiative transfer modeling of Class 0 and Class I sources and found that they were best explained by a mean radial profile with $n\sim1.6$. \citet{Enoch08} measured power law profiles for protostellar cores in Perseus, Serpens and Ophiuchus and deduced exponents of $n=1.3-1.5$. The radial profiles of our cores are therefore in good agreement with previous theories and observations, despite their asymmetrical nature.



We have considered the cores as if they were spheres in order to compare to the literature. However, it should be noted that for a perfectly cylindrical, non-magnetised filament, a constant density profile is expected along the major axis of the filament and a steep $\rho\propto x^{-4}$ density profile along the minor axis \citep{Ostriker64,Hennebelle03,Tilley03}. 

\begin{figure}
\begin{center}
\includegraphics[width=2.5in]{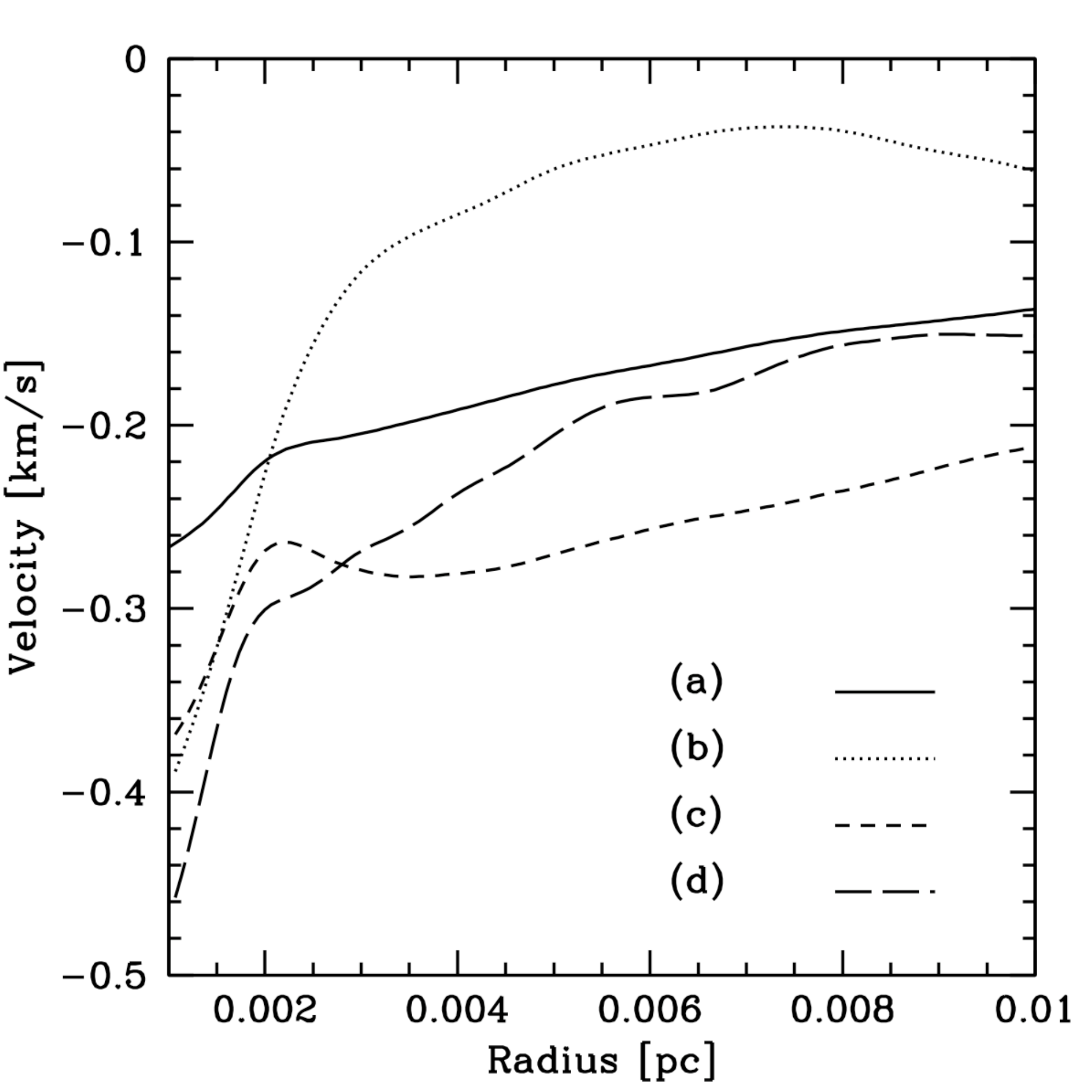}
\caption{Mass-averaged radial velocity profiles for cores (a)--(d). The infall is supersonic close to the sink and decreases outward. The sound speed for a molecular gas at $10$ K is of the order of $0.2$ km/s.}
\label{radialvel}
\end{center}
\end{figure}

The radial velocity profiles of the cores are shown in \fig~\ref{radialvel}. The central regions of all the cores are collapsing supersonically onto the sink particle. The sound speed for a molecular gas at $10$ K is of the order of $0.2$ km/s, meaning that the flow in the cores is only mildly supersonic. However, we are unable to accurately resolve scales of less than $0.001$ pc due to the sink particle. Moreover, the sudden steepening seen in the velocity profile at the centre may be due to the inability to accurately follow the pressure gradient accross the sink particle, which would lead to an artificial increase in the acceleration.  The central infall velocity predicted by the Larson-Penston model is $-3.3 c_s$. However, as \citet{Hunter77} has shown, this value is rarely approached apart from at the very central regions of numerical simulations. \citet{Ogino99} carried out simulations of collapsing profiles and found that the radial velocity profiles were in good agreement with the $ v \propto r^{-1/2}$ profile expected. Our profiles are not so simple, due to the more complex geometry, but like the spherical isothermal collapse models, they do collapse more quickly at the centre. 

\begin{figure*}
\begin{center}
\begin{tabular}{c c}
\includegraphics[width=2.5in]{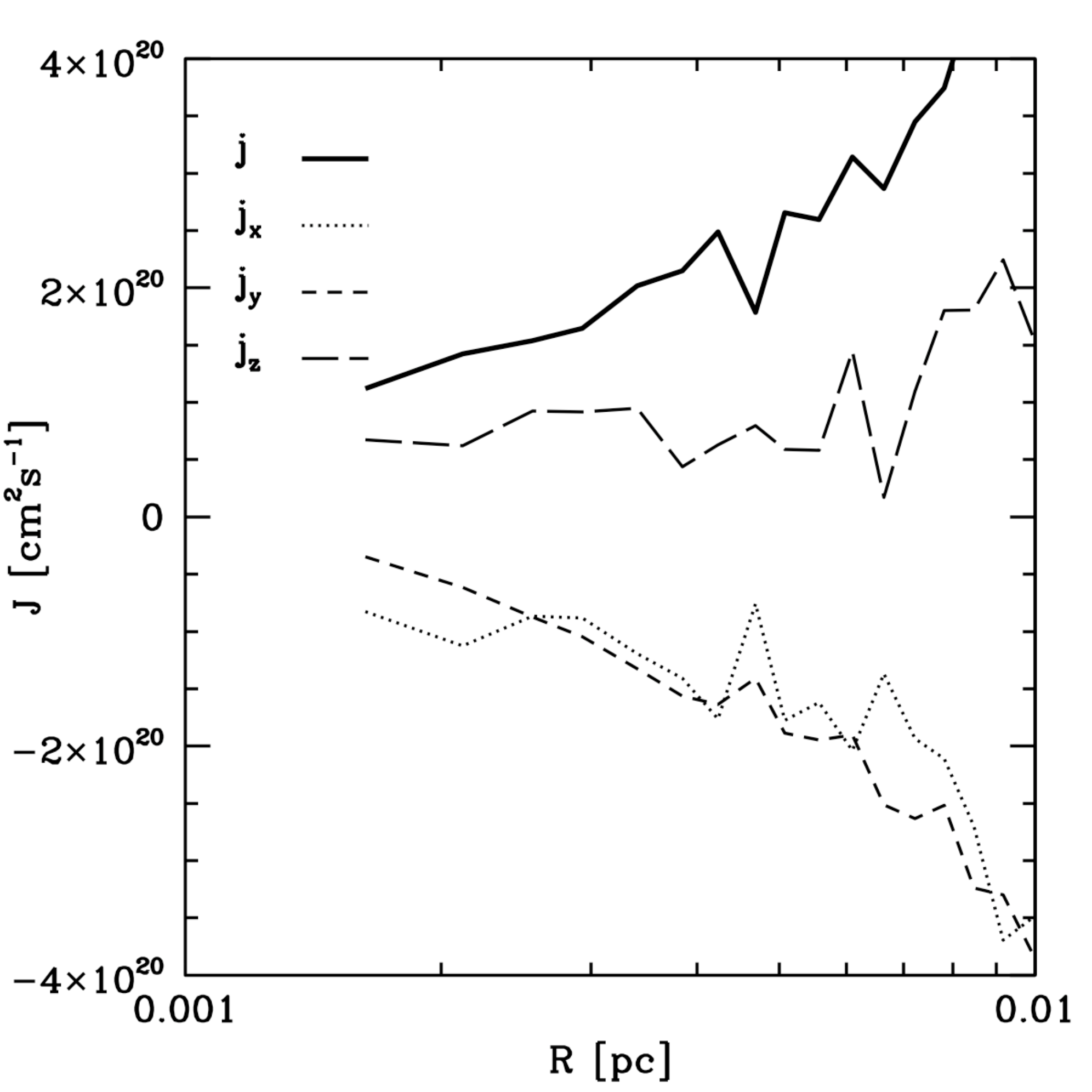}
\includegraphics[width=2.5in]{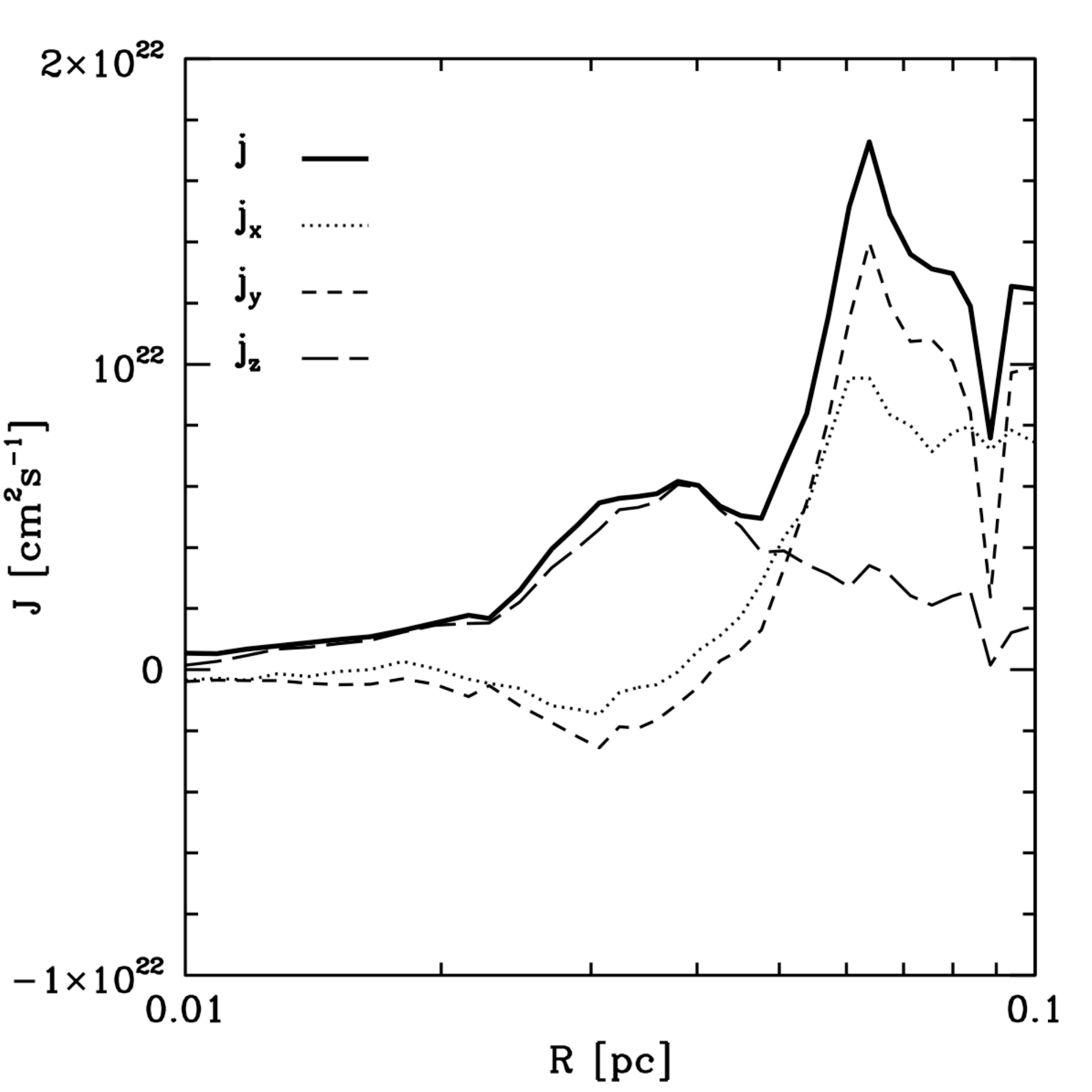}\\
\includegraphics[width=2.5in]{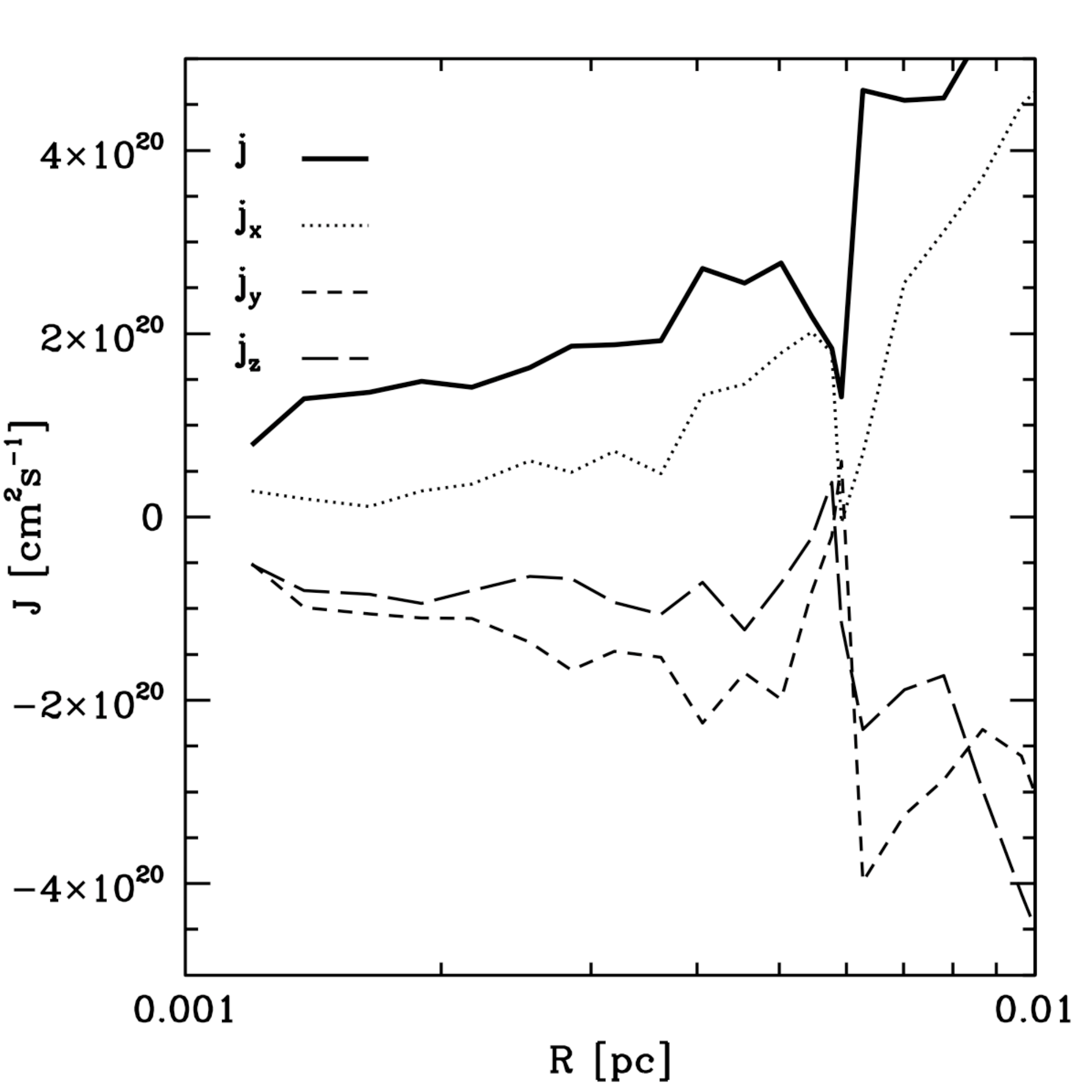}
\includegraphics[width=2.5in]{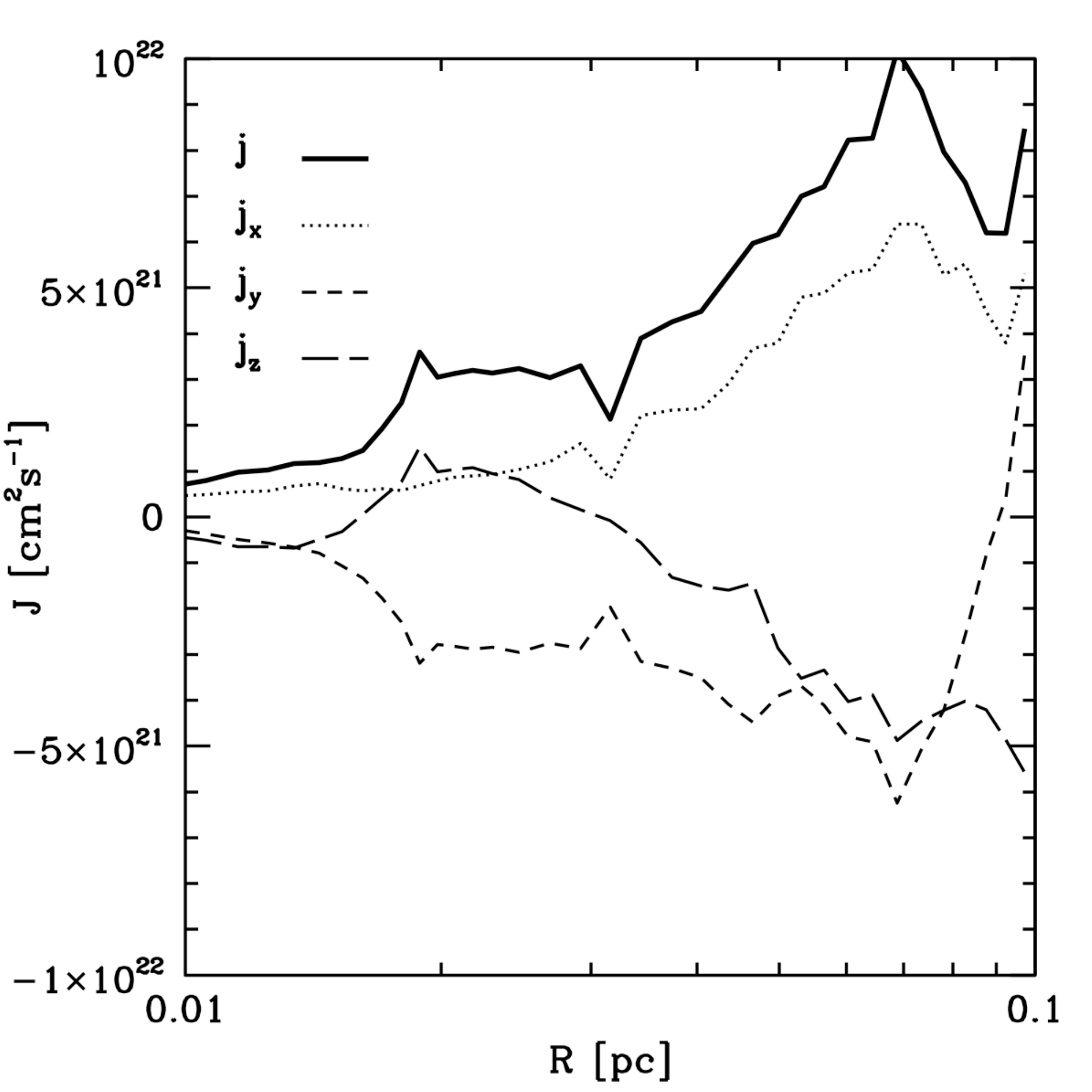}\\
\end{tabular}
\caption{Radial profiles of specific angular momentum for cores (b) \textit{top} and (d) \textit{bottom}. The left hand panels shows the specific angular momentum and its three components in the central core, and the right hand panels shows the specific angular momentum and its three components in the surrounding gas, out to a distance of $0.1$~pc. The relative magnitude of the angular momentum components alters radially, meaning that the angular momentum vector of the core will be constantly shifting as it accretes new material.}
\label{angmom}
\end{center}
\end{figure*}

A key factor determining the evolution of the cores is the evolution of their angular momentum. In \fig \ref{angmom} we show the angular momentum profiles of two of the cores and their environments. The relative magnitude of the three components of the angular momentum vector changes radially outward, meaning that the angular momentum vector of the core will shift as the core accretes new material. Disks formed in the centre of these cores will shift in orientation as new material is accreted. The specific angular momentum inside the cores is of the order of $10^{20}$ cm$^2$s$^{-1}$ and is smaller and more homologous than that outside ($ j \approx 10^{21} - 10^{22}$ cm$^2$s$^{-1}$), where the environment is supersonically turbulent and there are multiple structures. The specific angular momentum decreases radially towards the centre of the collapsing cores due to gravitational torques, as discussed by \citet{Jappsen04}. The observed angular momentum for the cores will therefore be sensitive to the physical size scale probed by the observations.

\begin{figure}
\begin{center}
\includegraphics[width=2.5in]{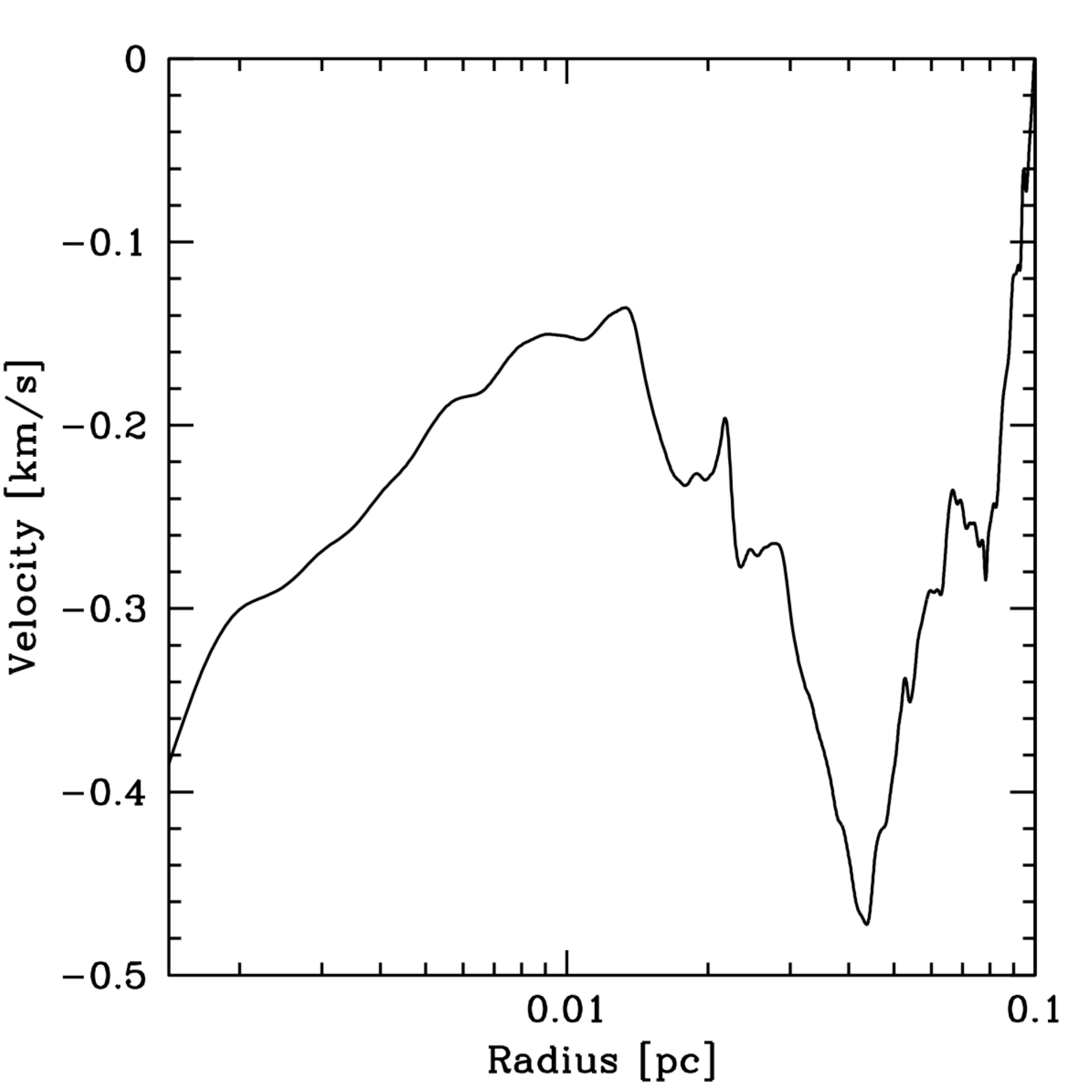}
\caption{Mass-averaged radial velocity profile of core (d). There are two stages of collapse affecting the core. On scales of $0.01$ pc there is the collapse of the core itself and on scales of $0.1$ pc there is the collapse of the region of gas within which the core is embedded.}
\label{radial64}
\end{center}
\end{figure}

\begin{figure}
\begin{center}
\includegraphics[width=2.5in]{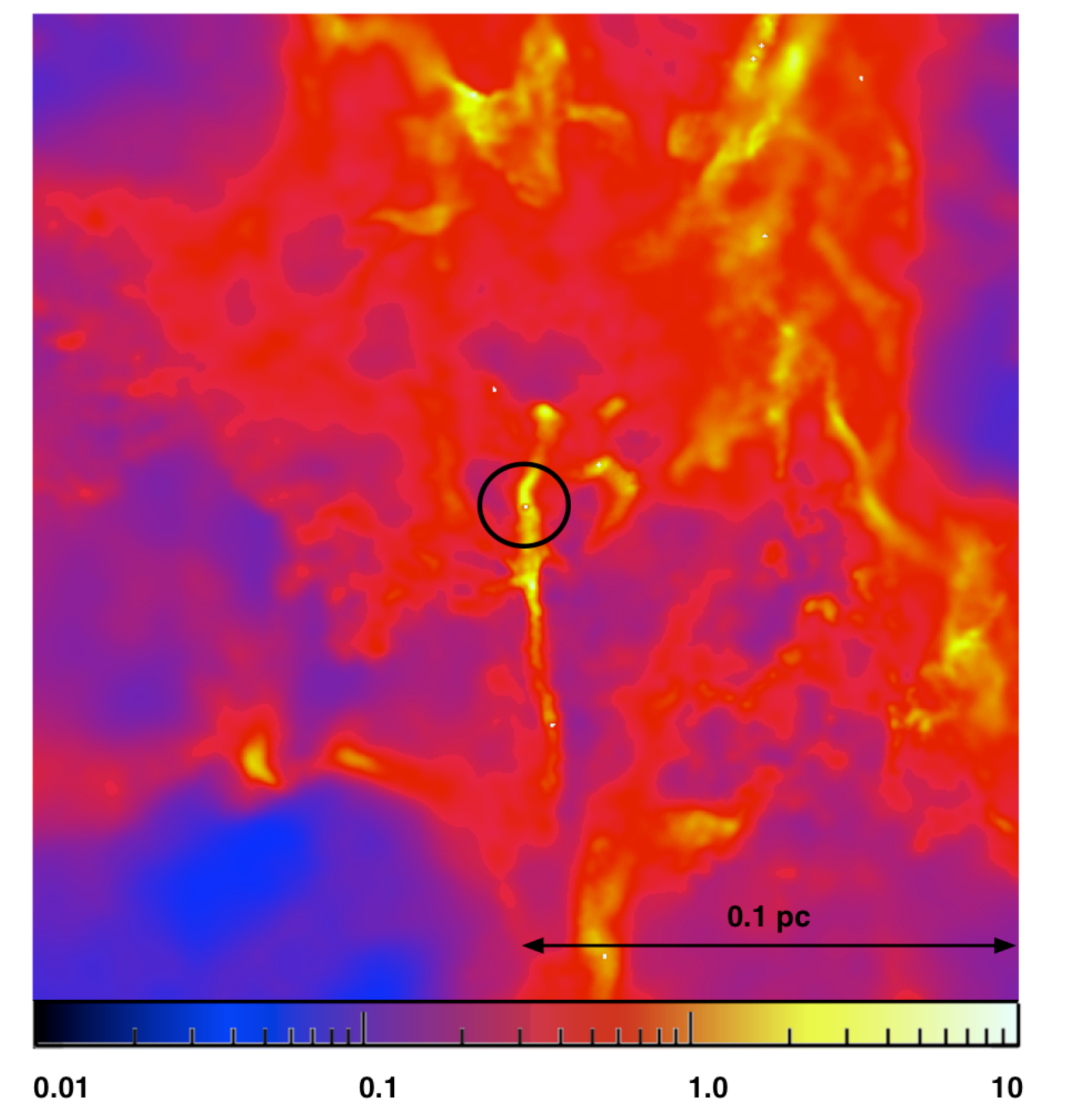}
\caption{Column density projection in \gcms of the local environment of the filamentary core (d), shown in the circle. There are multiple condensed structures within $0.1$pc of the core.}
\label{env64}
\end{center}
\end{figure}

As an illustrative example, Figures \ref{radial64} and \ref{env64} show the radial velocity and a column density projection for the wider environment of the filamentary core (d). In \fig \ref{radial64} there is a second strong signature of collapse at radii outside the core. This is a common occurence in our core sample, and represents the wider collapse of the region within which the core is located. This infalling material can also be accreted by the central sink, and in the case of massive stars constitutes the majority of the final sink mass (see SLB09). The geometry of the material involved in this secondary stage of accretion is discussed in Section \ref{sec:acc1}. This widespread infall is a clear indicator of the dynamic nature of the core's formation process, and it would not be seen in a singular isothermal sphere type collapse, where outside the rarefaction wave the gas is hydrostatic. \fig \ref{env64} shows a column density projection centred on core (d). The filament within which the core is situated is clearly visible and it is surrounded by a complex morphology of cores and filaments.

In conclusion, the radial behaviour of the collapsing cores is in good agreement with previous models. The core structure can therefore be thought of as a superposition of two effects. Firstly, there is the high density peak and steep radial decrease in mean density due to gravitational collapse. Secondly, there is the effect of the density field of the molecular cloud which is highly filamentary and irregular. It is this medium which has rapidly undergone collapse, and there is no time or mechanism for this structure to be homogenised \citep{Klessen00}. For example, the irregular cores seen by \citet{Stutz09} will have their non-axisymetric geometries enhanced during collapse, as denser regions collapse more swiftly than less dense ones.

\section{Accretion onto the core}
\subsection{Geometry of accreted gas}\label{sec:acc1}
Generally low mass and high mass sinks have different accretion histories. Typically low mass sinks only have accretion from within their core as outlined in the previous section. However, the growth of higher mass sinks is dominated by a secondary stage of accretion; the accretion of additional infalling material from outside the original core radius. We shall now examine whether this additional material is also inhomogeneously distributed.

\begin{figure*}
\begin{center}
\includegraphics[width=6in]{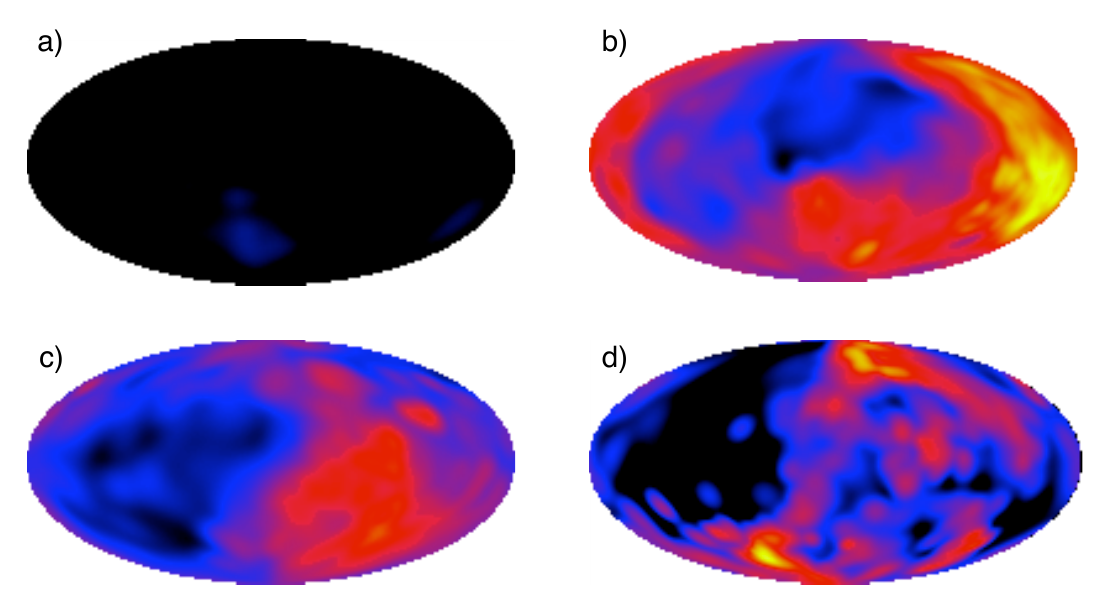}
\caption{The positions at which accreted material passes through a shell of radius $r=0.01$ pc around the sink over 20,000 yr. The amount accreted varies between the cores: the spherical core (a) accretes almost no additional material, but core (b) gains considerably. The accreted material comes from preferential directions and is not uniformly distributed.}
\label{CDaccinner}
\end{center}
\end{figure*}

\begin{figure*}
\begin{center}
\includegraphics[width=6in]{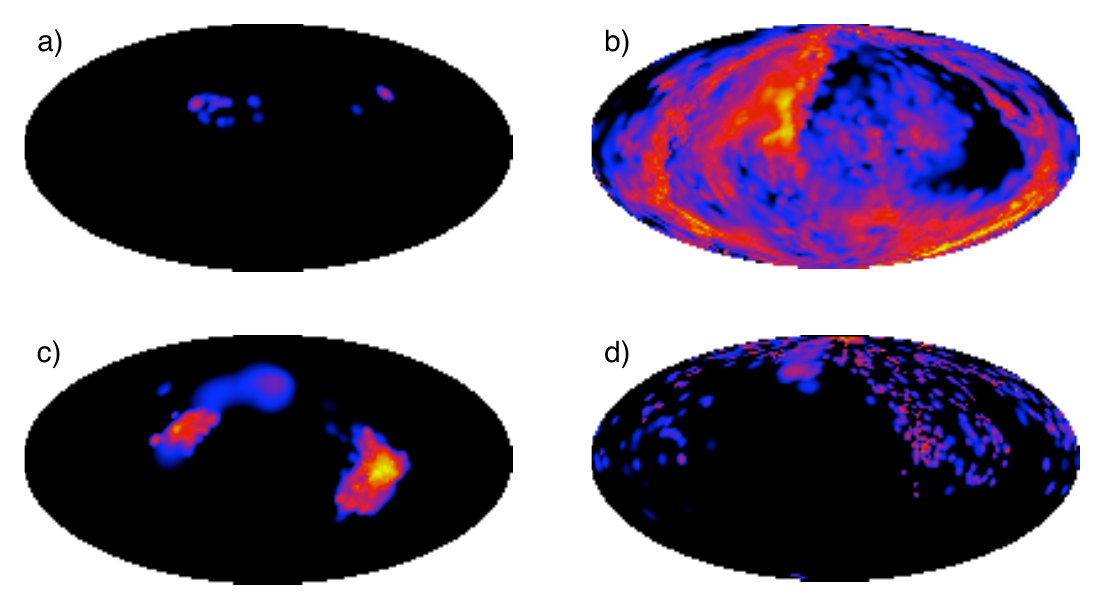}
\caption{The positions at which accreted material passes through a shell of radius $r=0.1$ pc around the sink over 20,000 yr. At this larger radius the accreted material is clearly coming from just a few directions. Core (c) is an obvious example of accretion purely along a filament.}
\label{CDaccouter}
\end{center}
\end{figure*}

To answer this question, we use a modification of the previous Hammer projection technique. As the SPH method is Lagrangian, it is possible to flag each individual gas particle that becomes accreted by a sink. We take a shell at the boundary of the core, and track the positions at which flagged particles from the external environment pass through the shell. We do this over a period of $2\E^4$ years (which is about the mean dynamical time of the bound pre-stellar cores found in SCB09) and then make Hammer projections of the integrated density on the shell in a similar manner to before. Figures \ref{CDaccinner} and \ref{CDaccouter} show the resulting `accretion surfaces' of material which passes through shells at $r=R_c=0.01$ pc and $r=0.1$ pc around the representative cores over the time interval. We do not specify the absolute values on this surface as it is merely a way of visualising the direction at which accreted material approaches the core. It is not a real density.

The cores in \fig \ref{CDaccinner} show differing amounts of additional accreted material passing through the core surface at $r=R_c$. For instance, core (a) has almost no additional accretion from its external environment, but core (b) has a large amount. The distribution of accreted material is again highly anisotropic, and shows a resemblance to the column density distribution in \fig \ref{CDprojections}. Accreted material from the environment is therefore entering the core along the high column density filaments.

\fig \ref{CDaccouter} shows how the accreted material enters the larger environment of the core at $r=0.1$ pc. The distributions of accreted material through the shells at $r=R_c$ and $r=0.1$ pc are different for several reasons. Firstly, most of the material passing through the shell at $r=0.1$ pc does not reach the inner shell in the time period considered here, and so is not included in \fig \ref{CDaccinner}. Secondly, the external molecular cloud structure is clumpy, with many voids that contain little gas, and which do not contribute to the accretion flow. Thirdly, the flow of material towards the cores is not necessarily purely radial. Instead, the accreted material often flows along the path of the high density filaments. \fig \ref{CDaccouter} shows that the additional accreted material is coming preferentially from just a few directions and large sections of the surface have no accreted material passing through them.

To get a more conceptual view of how the accretion is proceeding, let us examine the representative cores in more detail. Core (a) has very little accretion from its enviroment and becomes a low mass star ($0.63 \: {\rm M_{\odot}}$). Core (b) undergoes a significant amount of accretion from its environment and at the end of the simulation has a mass of $11.52 \: {\rm M_{\odot}}$. It has a large amount of material passing through the $r=0.1$ pc surface from a variety of directions. However, once again some directions contribute more than others. Core (c) is a clear example of accretion along a filament. Core (d) is also accreting along a filament, but most of the accreted material comes from close to the sink.

We carry out a similar analysis to the covariance function analysis used in Section \ref{sec:global} and classify the accretion surfaces of the entire core sample. In this case, we found it more useful to use the 
function $Q(d\theta)$ to classify the surfaces, rather than the covariance $cov(d\theta)$, as the `accretion density' that we are dealing with here is not a real quantity, and so it makes sense to use a dimensionless function. By normalising $Q(d\theta)$ by its maximum value, it is possible to make a fair comparison between the cores, regardless of how much they accrete. The class definitions are very similar to those used previously. For Type 0 cores, $Q(d\theta)$ remains almost flat, with deviations no larger than 20\%, while for Type 1 cores, it decreases monotonically as we move away from zero angular separation. For Type 2 cores, $Q(d\theta)$ shows a double-peaked structure. The sample of cores is smaller in this case, as the requirement of following the accretion for $2\E^4$ years means that cores which form close to the end of the simulation must be excluded. The accretion surfaces which the gas external to the core flows through are even more irregular than the column density projections, and almost all the cores have multiple angles from which they receive no additional material.

\begin{figure}
\begin{center}
\includegraphics[width=2.5in]{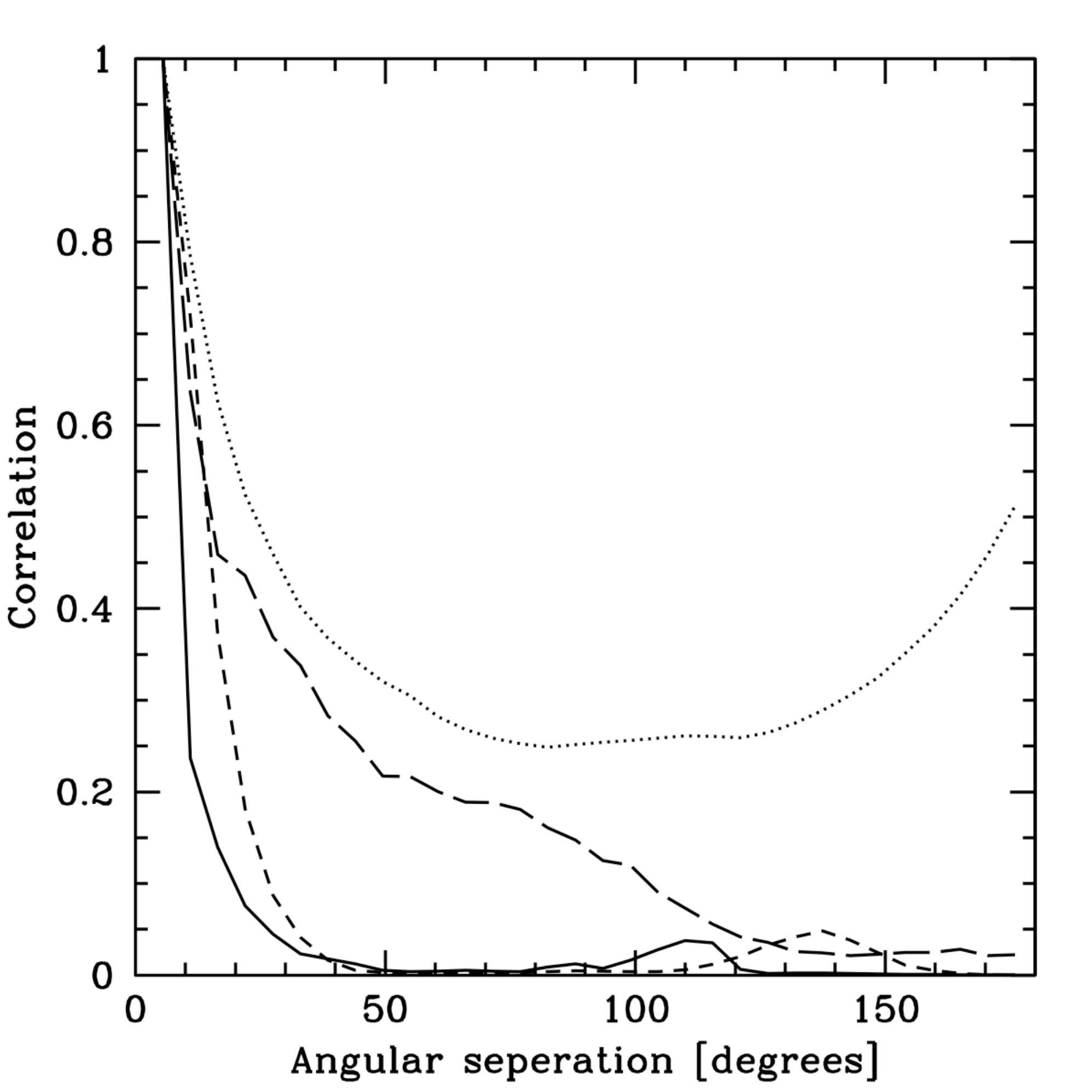}
\caption{Extent to which the accretion surfaces at $r=0.1$ pc of the four representative cores
are correlated, as quantified by $Q(d\theta)$. For most cores, we see only a single sharp peak,
corresponding to accreted material coming primarily from a single direction.}
\label{Corr_acc}
\end{center}
\end{figure}

\begin{table}
	\centering
	\caption{The number of cores assigned to each classification. The additional material accreted by the cores is even more inhomogeneously distributed than the gas in the cores themselves.}
		\begin{tabular}{c c c c c}
   	         \hline
	         \hline
	         Type & Number. & Number  & Percentage & Percentage\\
	         &R=$0.01$pc &R$=0.1$pc &R$=0.01$pc &R$=0.1$pc\\
	         \hline
	         0 & $3$ & $16$ & $1.3\%$ & $7.1\%$ \\
	         1 & $111$ & $170$ & $49.1 \%$& $75.2\%$\\
	         2 & $112$ & $40$ & $49.6 \%$ & $17.7\%$\\
		\hline
		\end{tabular}
	\label{type_acc}
\end{table}

\fig \ref{Corr_acc} shows $Q(d\theta)$ for the accretion surfaces of the representative cores at the $r=0.1$ pc shell. There are two key differences with the core column density covariance functions shown in \fig \ref{type_acc}. Firstly, the contrast is much larger than before. This is because there are areas of the accretion surface which are completely devoid of additional accreted gas, and hence produce zero correlation. Secondly, the functions are more irregular, owing to the irregular distribution of the accreted material. \tab \ref{type_acc} shows the classifications of the accretion surfaces for the entire core population. An almost negligible fraction of the cores have uniform, spherical accretion surfaces. The majority of the accreted material comes from just one or two directions, and in the case of the $r=0.1$ pc shell, it is almost all coming from just one direction.

In summary, the individual cores have different accretion behaviours, but they all share the property of non-axisymmetry. This is unsurprising, as it is well known that the larger-scale structure in molecular clouds is complex and filamentary, and as the accreted material is drawn from this distribution, the accretion flows onto the sink must also be irregular. Accretion onto cores from the external molecular cloud environment is primarily along the filaments in which the cores are embedded. Therefore the angular momentum being added to the protostellar disks from the environment is typically unbalanced.

\subsection{Final sink masses}

\begin{figure*}
\begin{center}
\begin{tabular}{c c}
\includegraphics[width=2.5in]{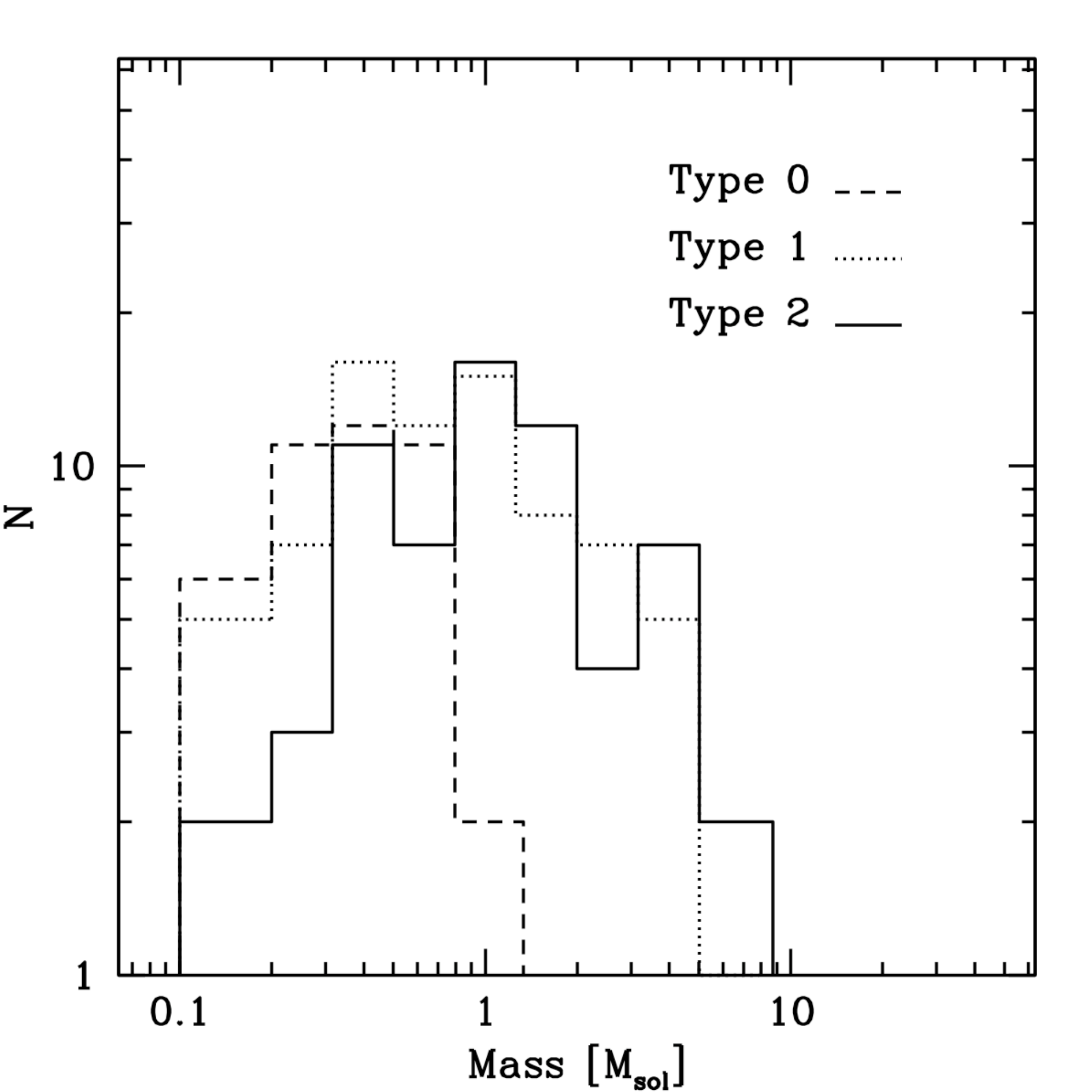}
\includegraphics[width=2.5in]{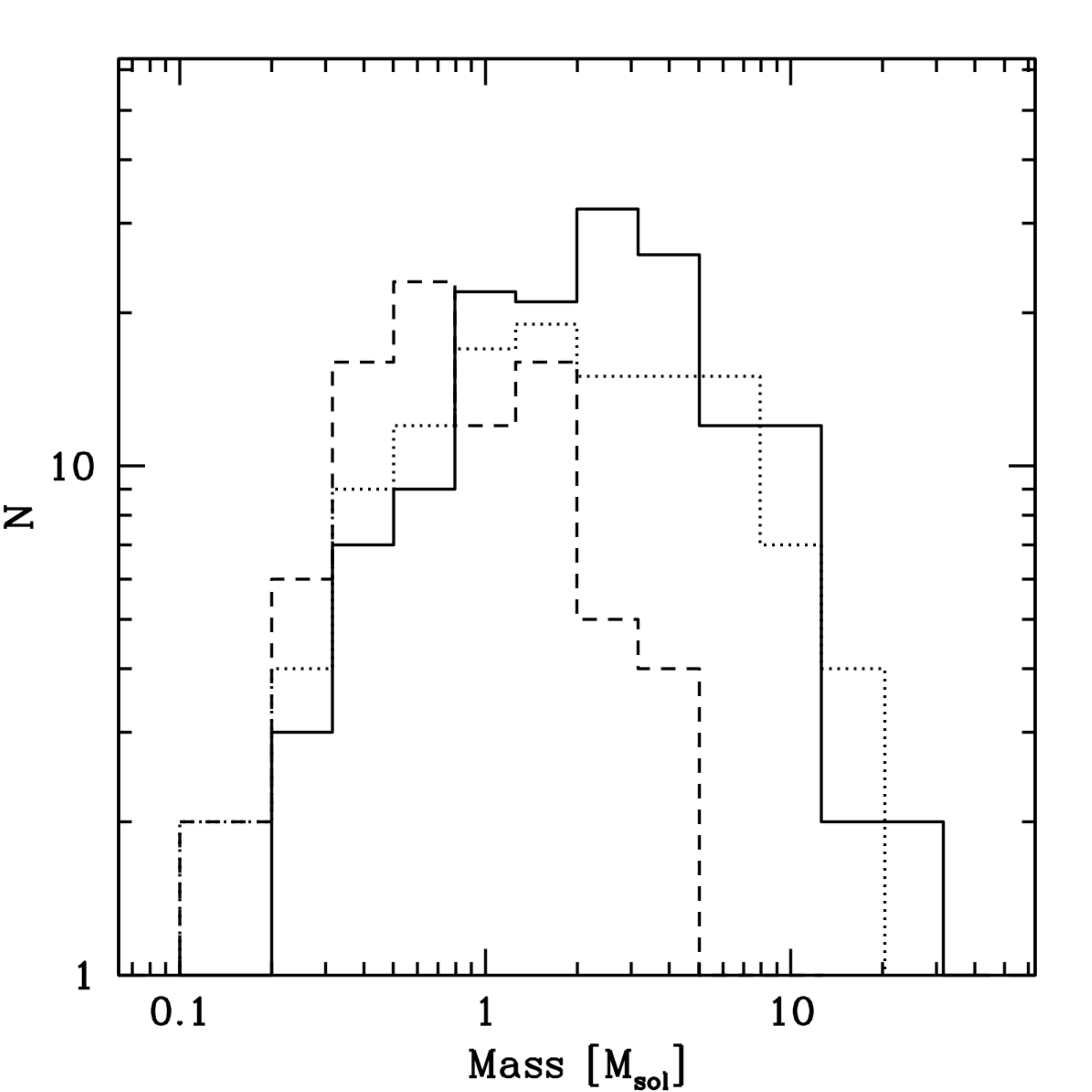}\\
\end{tabular}
\caption{The distribution of final central sink masses formed in each type of core after \textit{left} one simulation dynamical time ($4.7\E^5$ yr ) and \textit{right} at the end of the simulation ($6.6\E^5$ yr). Sinks within Type 2 cores are the most successful at gaining additional mass from their wider environment.}
\label{Hmass}
\end{center}
\end{figure*}

The final outcome of the accretion process also varies between the types of cores. \fig \ref{Hmass} shows histograms of the masses of the central sinks formed from each type of core at two snapshots in time. The right hand panel shows the sink masses at one dynamical time ($4.7\E^5$ yr ) and the left hand panel at the end of the simulation ($6.6\E^5$ yr). The thermal feedback in this simulation is unable to prevent accretion (see future papers for a full discussion) so we cannot ascertain the final sink mass reliably. However it is clear that the filamentary cores have been more successful at gaining additional mass from outside the core region. The entirety of the high mass end of the sink mass distribution in \fig \ref{Hmass} is formed from Type 1 or Type 2 cores.

There are two possible reasons for this difference in sink masses. Firstly, the filamentary cores are predominantly formed in denser, more clustered regions in which there is more gas outside the core available for accretion. Secondly, as shown in the Section \ref{sec:acc1}, the Type 1 and 2 cores are part of larger external filamentary accretion streams along which additional mass is channeled towards the sink. The filamentary nature of molecular clouds therefore contributes directly to the assembly of stellar masses.


\section{Discussion}

We have classified the surroundings of sink particles in an SPH simulation of a giant molecular cloud, and discovered that in three quarters of cases the cores depart significantly from spherical symmetry.  Furthermore, accretion onto these cores from the environment is even more asymmetric and takes place primarily along filaments. Traditionally, spherical models have been used to describe star formation, and so in this section we qualitatively discuss some of the implications of non-axisymmetry.

\subsection{Detecting Infall}

The classical signature of collapse is a double-peaked line profile which is shifted to the blue \citep{Evans99}. This is what would be expected from a spherical cloud undergoing inside-out collapse. A static envelope produces the main self-absorption dip in the line, the back of the cloud collapsing towards the observer produces the blue peak, and the front of the cloud collapsing away from the observer produces the red peak. Our Type 0 cores are close to spherical and should have line profiles close to this expected result.

Now consider filamentary geometry, as in the case of Type 2 cores. Along a collapsing filament, the same arguments hold, and the expected double-peaked line profile should be produced. However, when perpendicular to the filament, the line profile may differ. Moreover, for the case of the Type 1 cores, where there is only one strong over-density, the line profile may entirely lack a strong blue or red peak. 

Therefore, from simple dimensional arguments we can estimate how common departures from a blue-skewed, double-peaked line profile will be. About one third of the cores are Type 1 ($33.7\%$), and $42.1\%$ of the cores are Type 2, and for the latter there is a roughly two in three chance that we will
be looking perpendicular to the filament axis. Therefore, conservatively, we would expect $0.337+ 2/3 \times 0.421 \simeq 60 \%$ of cores which are collapsing to depart from the standard line profile. We hope to investigate this more fully in future work, in which we will use radiative transfer modeling of our cores to examine how line profiles change with viewing angle.

In surveys of both protostellar and prestellar cores, there is an overabundance of blue line asymmetries over red asymmetries, which suggests that most objects are characterised by collapse. However, there are relatively few unambiguous examples of the expected blue-shifted line profiles \citep[e.g.][]{Gregersen97,Lee99,Andre07,Chen10}. The high fraction of filamentary cores seen in this study provides a possible explanation for this.

\subsection{Preventing Accretion}

At present, it is hard to carry out detailed radiative transfer modeling within clustered cores, and generally cores are modeled individually. The Bonnor-Ebert sphere, perhaps with some turbulence injected, is a traditional choice of initial conditions. However, our collapsing cores depart significantly from this geometry, which may have sizable implications for the final results. 

In the case of massive stars, there are several current efforts to model radiative effects on core environments \citep[e.g.][]{Yorke02,Krumholz09, Kuiper10,Peters10}. \citet{Kahn74} showed that for spherically symmetric accretion, radiation pressure will halt accretion before truly massive stars can form. Recently \citet{Krumholz09} showed in an initially spherically symmetric calculation that instabilities on the surface of a radiation-filled bubble surrounding the protostar create self-shielding filaments of gas which continue to channel material onto the disk. Similar results are seen when ionizing radiation is included in the models \citep{Peters10}. The cores in our simulation (which are the precursors of both low and high mass stars) contain dense filaments and streams from their first formation, and therefore accretion can continue unimpeded throughout. Generally, the geometries of the collapsing cores studied here encourage the escape of radiation, as most of the core surface area is at low column densities, which radiation can pass through easily. The radiation is likely to simply pass around the concentrated filaments of high column density, high inward pressure gas, allowing accretion to proceed unimpeded through these filaments.

The accretion onto the core from the external molecular cloud environment is also through irregular filaments. Again, this will make it harder to shut off accretion as a uniform radiation field will likely prove too weak to stop the densely collimated inflows. \citet{Wang10} recently modeled the effects of feedback from protostellar jets and found that these could possibly reduce the accretion through filaments. 

\subsection{Binary Formation and Angular Momentum}

The irregular inflows seen here have implications for the stability of the collapsing cores. Even small departures from spherical symmetry make cores more liable to fragment \citep{Bonnell92,Burkert93,Bodenheimer95}. Moreover, the flow of material onto the disk will be irregular, which increases the likelihood of disk fragmentation and encourages binary formation. In studies of core fragmentation with the effects of ideal MHD included \citep{Hennebelle08d,Hennebelle08b} there is some difficulty in producing binaries at all. These authors suggest that one way to overcome this barrier would be if the angular momentum transferred to the core was unbalanced, which would encourage fragmentation. The angular momentum vector of the gas surrounding our sink particle varies in direction and magnitude and therefore the angular momentum of the core will be constantly changing as it collapses \citep{Jappsen04}.

\subsection{Outstanding Issues}

This work considers hydrodynamic collapse and neglects magnetic fields. Magnetic pressure slows gravitational collapse \citep{Strittmatter66,Mouschovias76,Heitsch01}, and therefore there may be more time for structures to become homogenised. However, uniform magnetic fields also introduce a preferred axis of collapse, which would exacerbate the anisotropies in the collapsing cores. It is unclear which of these effects will be more significant, or whether they would to some extent cancel out.

Magnetohydorodynamic (MHD) calculations of molecular clouds have studied the morphologies of clumps and cores with magnetic fields \citep{Li04,Tilley07}, and found they were also filamentary and irregular. This suggests that the addition of molecular fields would not substantially alter our results. MHD codes are grid based, and require adaptive mesh refinement in order to match the dynamical range of SPH, which enables us to resolve scales of $10$ pc down to a few hundred AU. It is this property which has allowed us to both generate a large dataset of cores, and to probe their structure on the smallest scales. There have been valiant attempts to introduce magnetic fields into SPH \citep{Price08,Price10}, but as yet there are no large scale simulations of an entire molecular cloud, as would be needed to robustly assemble enough cores for a statistical analysis.

Our simulation also lacks full radiative transfer, and instead uses a crude heating prescription. This is almost certainly wrong, particularly as it assumes spherical symmetry, which we have just shown to be a poor assumption. However, this heating prescription, if anything, overestimates the heating from the protostars, and therefore is likely to support the core and smooth its structure. Without this prescription, our cores would be even more filamentary. We include it in this analysis as there will be a supporting force from the radiative transfer, and it is better to overestimate it than underestimate it.

\section{Conclusions}
We have classified the structures of Class 0 cores in a large SPH simulation of a giant molecular cloud.  We define the cores as the material within $r=0.01$ pc of the sink when it first forms. Our main conclusions are as follows.
\begin{enumerate}
\item The density distribution in collapsing cores is highly non-axisymetric.
\item Three-quarters of cores are better described by filamentary geometries than by spherical ones.
\item The density profiles decrease radially according to a power law, and are consistent with observations and previous studies of collapse. 
\item The angular momentum vector of a core and its surrounding gas varies with radius, and so the inclination of protostellar disks formed within the cores will alter with time.
\item The cores have mildly supersonic infall velocities at their centre.
\item Accretion onto the central protostar can be broken into two phases. First the original core is accreted, and then, in the case of higher mass stars, infalling material from outside the original core radius is accreted.
\item Additional accreted material from the wider molecular cloud environment primarily flows towards the cores along the dense filaments in which the cores are embedded.
\item Protostars formed in filamentary cores obtain a higher final mass from accretion.
\end{enumerate}
Due to the non-axisymetric nature of the cores, the line profiles of collapsing cores are likely to be more complex than spherical models suggest. It will be harder to slow down, or prevent core accretion through feedback, as most of the surface of the cores is at low column densities, allowing radiation to escape \citep{Dale08}, but a significant fraction of the infalling mass has high column density and inward pressure. The asymmetric nature of the cores will also make them more vulnerable to fragmentation and binary formation.

\section*{Acknowledgements}
We thank Amy Stutz, John Tobin, Javier Ballesteros-Paredes, Robi Banerjee, Philipp Girichidis, Patrick Hennebelle, and Enrique Vazquez-Semadeni for many stimulating discussion on cloud shapes and morphologies. We also thank an anonymous referee for their feedback.

R.S.K., S.C.O.G., and P.C.C.\ acknowledge financial support from the {\em Landesstiftung Baden-W{\"u}rttemberg} via their program International Collaboration II (grant P-LS-SPII/18) and from the German {\em Bundesministerium f\"{u}r Bildung und Forschung} via the ASTRONET project STAR FORMAT (grant 05A09VHA). The authors furthermore gives thanks for subsidies from the {\em Deutsche Forschungsgemeinschaft} (DFG) under grants no.\ KL 1358/1, KL 1358/4, KL 1359/5, KL 1358/10, and KL 1358/11, as well as from a Frontier grant of Heidelberg University sponsored by the German Excellence Initiative. The work presented in this paper was assisted by the European Commission FP6 Marie Curie
RTN CONSTELLATION (MRTN-CT-2006-035890). This work was supported in part by the U.S. Department of
Energy contract no. DE-AC-02-76SF00515. R.S.K. also thanks the KIPAC at Stanford University and the Department of Astronomy and Astrophysics at the University of California at Santa Cruz for their warm hospitality during a sabbatical stay in spring 2010.
\bibliography{Bibliography}
\label{lastpage}

\end{document}